\documentclass[11pt,a4paper,twoside]{article} 
\usepackage{fmrep}
\usepackage{amsmath,amssymb,amsfonts}
\usepackage{graphicx}
\usepackage{enumitem}
\usepackage{algpseudocode}
\usepackage{rotating}
\usepackage{multirow}
\usepackage{caption}
\usepackage[frozencache]{minted}
\usepackage{booktabs}
\usepackage{color}
\usepackage{float}
\usepackage{listings}
\usepackage{changepage}
\usepackage[T1]{fontenc}
\usepackage{lmodern}
\usepackage{sectsty}
\usepackage{titlesec}
\usepackage{xcolor}
\usepackage{pagecolor}
\usepackage{sourcecodepro}
\usepackage{afterpage}
\definecolor{bluep}{rgb}{0.2, 0.2, 0.6}
\definecolor{fgreen}{rgb}{0.0, 0.27, 0.13}
\definecolor{maroon}{rgb}{0.76, 0.13, 0.28}
\usepackage[ruled,vlined]{algorithm2e}
\usepackage{placeins}
\usepackage{pdfpages}
\providecommand{\keywords}[1]{\textbf{\textit{Keywords:}} #1}
\renewcommand{\theequation}{\arabic{equation}}
\newenvironment{longlisting}{\captionsetup{type=listing}}{}
\FLD{01}{2021}
\title{On the performance of GPU accelerated q-LSKUM based meshfree solvers in Fortran, C++, Python, and Julia }
\shorttitle{GPU Accelerated Meshfree Solvers in Fortran, C++, Python, and Julia }
\author{Nischay Ram Mamidi, Kumar Prasun, Dhruv Saxena, Anil Nemili \affil{ Department of Mathematics\\
BITS Pilani - Hyderabad Campus \\ Hyderabad 500078, India \\{\em Email:} {\tt {nischay,f20150845,f20191369,anil}@hyderabad.bits-pilani.ac.in }}
Bharatkumar Sharma \affil{NVIDIA \\ {\em Email:} {\tt bharatk@nvidia.com}}
SM Deshpande\affil{524, Tata Nagar, Bengaluru, India \\
{\em Email:} {\tt desh1942@gmail.com }}}
\shortauthor{Nischay, Kumar, Dhruv, Anil, Bharat and Deshpande}
\date{May 19, 2021 \\
Revised May 19, 2021}

\begin{document} 

%
\begin{titlepage}
\definecolor{titlepage-color}{HTML}{fdeed4}
\newpagecolor{titlepage-color}\afterpage{\restorepagecolor}
\newcommand{\colorRule}[3][black]{\textcolor[HTML]{#1}{\rule{#2}{#3}}}
\begin{flushleft}
\noindent
\\[-1em]
\color[HTML]{000000}
\makebox[0pt][l]{\colorRule[000000]{1.3\textwidth}{2pt}}
\par
\noindent

{
  \vfill
  \noindent {\fontfamily{SourceSansPro-LF}\fontsize{16}{15}\selectfont \textbf{On the performance of GPU accelerated q-LSKUM based \\ \vspace{1mm} meshfree solvers in Fortran, C++, Python, and Julia}}
    \vskip 1em
  {\fontfamily{SourceSansPro-LF}\fontsize{14}{16}\selectfont \textbf{Scientific Computing Report - 01:2021}}
    \vskip 2em
   \noindent {\fontfamily{SourceSansPro-LF}\fontsize{12}{15}\selectfont \textbf{M. Nischay, P. Kumar, S. Dhruv, N. Anil, S. Bharatkumar, and S.M. Deshpande}}
  \vfill
   \includegraphics[height=1.8cm]{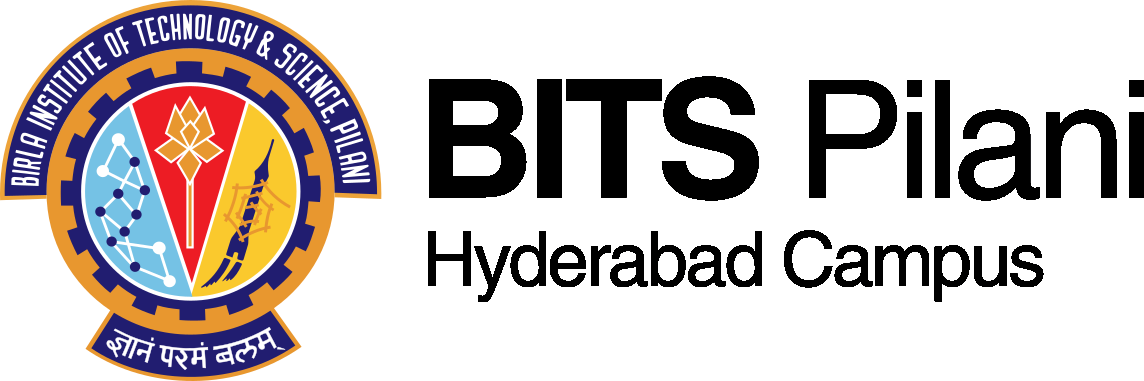}
}

\end{flushleft}
\end{titlepage}
%


\maketitle

%
\setcounter{page}{1}
\thispagestyle{empty}
\begin{abstract}
This report presents a comprehensive analysis of the performance of GPU accelerated meshfree CFD solvers for two-dimensional compressible flows in {\tt Fortran}, {\tt C++}, {\tt Python}, and {\tt Julia}. The programming model CUDA is used to develop the GPU codes. The meshfree solver is based on the least squares kinetic upwind method with entropy variables ($\boldsymbol{q}$-LSKUM). To assess the computational efficiency of the GPU solvers and to compare their relative performance, benchmark calculations are performed on seven levels of point distribution. To analyse the difference in their run-times, the computationally intensive kernel is profiled. Various performance metrics are investigated from the profiled data to determine the cause of observed variation in run-times. To address some of the performance related issues, various optimisation strategies are employed. The optimised GPU codes are compared with the naive codes, and conclusions are drawn from their performance. 
\end{abstract}
\keywords{{\tt Fortran}, {\tt C++}, {\tt Python}, {\tt Julia}, GPUs, CUDA, LSKUM, Meshfree methods.}
\section{Introduction}
High performance computing (HPC) plays a critical part in the numerical simulation of many complex aerodynamic configurations. Typically, such simulations require solving the governing Euler or Navier-Stokes equations on fine grids ranging from a few million to several billion grid points. To perform such computationally intensive calculations, the computational fluid dynamics (CFD) codes use either only CPUs or CPU-GPUs. However, for computations on multiple GPUs, CPUs tackle control instructions and file input-output operations, while GPUs perform the compute intensive floating point arithmetic. Over the years, GPUs have evolved as a competitive alternative to CPUs in terms of better performance, cost and energy efficiency. Furthermore, they consistently outperform CPUs in single instruction multiple data (SIMD) scenarious. \\ \\
In general, the GPU codes for CFD applications are written in traditional languages such as {\tt Fortran} or {\tt C++} \cite{hifiles-gpu-code,gpu-rans-jude-baeder,giles-hydra-gpu-code,fun3d-gpu}. In recent years, modern languages such as {\tt Python} \cite{python}, {\tt Julia} \cite{CUDAnative.jl-2018}, {\tt Regent} \cite{Regent15}, {\tt Chapel} \cite{chapel} have steadily risen in the domain of scientific computing. These languages are known to be architecture independent with the added advantage of easy code maintenance and readability. Recently, {\tt Python} has been used to develop an industry standard CFD code, {\tt PyFR} for compressible flows  \cite{pyfr}. 
To the best of our knowledge, a rigorous investigation and comparison of the performance of GPU codes for CFD written in both traditional and modern languages are not yet pursued. \\ \\
Towards this objective, this report presents a comprehensive analysis of the performance of GPU accelerated CFD solvers in {\tt Fortran}, {\tt C++}, {\tt Python}, and {\tt Julia}. The programming model CUDA is used to construct the GPU solvers. Here, the CFD solver is based on the meshfree Least Squares Kinetic Upwind Method (LSKUM) \cite{lskum-ghosh-aiaa}. The LSKUM based CFD codes are being used in the National Aerospace Laboratories and the Defence Research and Development Laboratory, India, to compute flows around aircraft and flight vehicles \cite{kobc,anandh-lskum-sepdynamic-jaircraft,anandh-lskum-viscous-aiaa-journal}. \\ \\
%
%
This report is organised as follows. Section \ref{sec-q-lskum} describes the basic theory of the meshfree solver based on $\boldsymbol{q}$-LSKUM. Section \ref{gpu-solver} presents the pseudo-code of the serial and GPU accelerated meshfree solvers. Section \ref{sec-naive-gpu-codes-performance} shows the performance of the naive GPU codes. A detailed analysis of various performance metrics of the kernels is presented. Section \ref{sec-opt-gpu-codes-performance} presents various optimistion strategies employed to enhance the computational efficiency. Furthermore, numerical results are presented to compare the performance of optimised GPU codes with the naive codes. Finally, Section \ref{sec-conclusions} presents the conclusions and a plan for future work. 
\section{Basic theory of q-LSKUM}
\label{sec-q-lskum}
The Least Squares Kinetic Upwind Method (LSKUM) \cite{lskum-ghosh-aiaa} belongs to the family of kinetic theory based upwind schemes for the numerical solution of Euler or Navier-Stokes equations that govern the compressible fluid flows. These schemes are based on the moment method strategy \cite{kfvs}, where an upwind scheme is first developed at the Boltzmann level. After taking appropriate moments, we arrive at an upwind scheme for the governing conservation laws. LSKUM requires a distribution of points, which can be structured or unstructured. The point distributions can be obtained from simple or chimera point generation algorithms, quadtree, or even advancing front methods \cite{lskum-smd-dsj}. \\ \\
 This section presents the basic theory of LSKUM for two-dimensional $\left( 2D \right)$ Euler equations that govern the inviscid compressible fluid flows. In the differential form, the governing equations are given by
\begin{equation}
\frac{\partial \boldsymbol{U}}{\partial t} + \frac{\partial \boldsymbol{Gx}}{\partial x} + \frac{\partial \boldsymbol{Gy}}{\partial y}  = 0
\label{ee-conservation-laws}
\end{equation}
Here, $\boldsymbol{U}$ is the conserved vector, $\boldsymbol{Gx}$ and $\boldsymbol{Gy}$ are the flux vectors along the coordinate directions $x$ and $y$, respectively. These vectors are given by
\begin{equation}
    \boldsymbol{U} = \begin{bmatrix}
    \rho \\ \rho u_1 \\ \rho u_2 \\ \rho e
    \end{bmatrix} \; \; \; \; 
    \boldsymbol{Gx} = \begin{bmatrix}
    \rho u_1 \\ p + \rho u_1^2 \\ \rho u_1 u_2 \\  \left( p + \rho e \right) u_1 
    \end{bmatrix} \; \; \; \;
    \boldsymbol{Gy} = \begin{bmatrix}
    \rho u_2 \\ \rho u_1 u_2 \\ p + \rho u_2^2 \\ \left( p + \rho e \right) u_2 
    \end{bmatrix}
\end{equation}
Here, $\rho$ is the fluid density, $u_1$ and $u_2$ are the Cartesian components of the fluid velocity along the coordinate directions $x$ and $y$, respectively. $p$ is the pressure, and $e$ is the specific total energy per unit mass, given by 
\begin{equation}
    e = \frac{p}{\rho \left(\gamma - 1\right)} + \frac{u_1^2 + u_2^2}{2}
\end{equation}
where, $\gamma$ is the ratio of specific heat of a gas. The conservation laws in eq. (\ref{ee-conservation-laws}) can be obtained by taking $\boldsymbol{\Psi}$ - moments of the $2D$ Boltzmann equation in the Euler limit \cite{kfvs}. In the inner product form, these equations can be related as 
\begin{equation}
\frac{\partial \boldsymbol{U}}{\partial t} + \frac{\partial \boldsymbol{Gx}}{\partial x} + \frac{\partial \boldsymbol{Gy}}{\partial y} = \left< \boldsymbol{\Psi}, \frac{\partial F}{\partial t} + v_1 \frac{\partial F}{\partial x} + v_2 \frac{\partial F}{\partial y} \right> = 0
\label{ee-be-me}
\end{equation}
Here, $F$ is the Maxwellian velocity distribution function, given by
\begin{equation}
    F = \frac{\rho }{I_0} \sqrt{\frac{\beta}{\pi}}  \left[ -\beta \left\lbrace \left(v_1 - u_1 \right)^2 + \left(v_2 - u_2 \right)^2 \right\rbrace - \frac{I}{I_0} \right] 
\end{equation}
where $\beta = 1/\left(2RT\right)$, $R$ is the gas constant per unit mass, $T$ is the absolute temperature, and $v_1$ and $v_2$ are the molecular velocities along the coordinate directions $x$ and $y$, respectively. $I$ is the internal energy, and $I_0$ is the internal energy due to non-translational degrees of freedom. In $2D$, $I_0$ is given by $I_0 = \left(2 - \gamma \right)/\left(\gamma - 1 \right) $. $\boldsymbol{\Psi}$ is the moment function vector, defined by 
\begin{equation}
    \boldsymbol{\Psi} = \begin{bmatrix}
    1, \; v_1, \; v_2, \; I + \frac{1}{2} \left(v_1^2 + v_2^2\right)
    \end{bmatrix}
\end{equation}
The inner products $ \left< \boldsymbol{\Psi}, F \right > $, $ \left< \boldsymbol{\Psi}, v_1F \right > $ and $ \left< \boldsymbol{\Psi}, v_2F \right > $ are defined as 
\begin{equation}
\left< \boldsymbol{\Psi}, F \right > = \int\limits_{\mathbb{R}^+ \times \mathbb{R}^2} \boldsymbol{\Psi} F dv_1 dv_2dI, \; \; 
\left< \boldsymbol{\Psi}, v_iF \right > = \int\limits_{\mathbb{R}^+ \times \mathbb{R}^2} \boldsymbol{\Psi} v_iF dv_1 dv_2 4dI, \; i = 1,2.
\end{equation}
Using Courant-Issacson-Rees (CIR) splitting \cite{cir-splitting} of molecular velocities, an upwind scheme for the Boltzmann equation in eq. (\ref{ee-be-me}) can be constructed as
\begin{equation}
\frac{\partial F}{\partial t } + v_1^{+} \frac{\partial F}{\partial x } +  v_1^{-}  \frac{\partial F}{\partial x } +  v_2^{+}  \frac{\partial F}{\partial y } + v_2^{-}  \frac{\partial F}{\partial y } = 0
\label{kfvs-3d-be}
\end{equation}
where, the split velocities $v_1^{\pm}$ and $v_2^{\pm}$ are defined as 
\begin{equation}
v_i^{\pm} = \frac{v_i \pm \left| v_i \right | }{2}, \; \; i = 1, 2. 
\end{equation}
The basic idea of LSKUM is to first obtain discrete approximations to the spatial derivatives in eq. (\ref{kfvs-3d-be}) using least squares principle. Later, $\boldsymbol{\Psi}$-moments are taken to get the meshfree numerical scheme for the conservation laws in eq. (\ref{ee-conservation-laws}). We illustrate this approach to determine the partial derivatives $F_x$ and $F_y$ at a point $P_0$ using the data at its neighbours. The set of neighbours, also known as the stencil of $P_0$, is denoted by $N\left(P_0\right)$ and defined by $N \left(P_0 \right) = \left\lbrace P_i : d\left( P_i, P_0 \right) < \epsilon \right\rbrace$. Here, $d \left( P_i, P_0 \right) $ is the Euclidean distance between the points $P_i$ and $P_0$. $\epsilon$ is the user-defined characteristic linear dimension of $N\left(P_0 \right)$.  \\ \\
To derive the least squares approximation of $F_x$ and $F_y$, consider the Taylor series expansion of $F$ up to linear terms at a neighbour point $P_i$ around $P_0$,
\begin{equation}
\Delta F_i = \Delta x_i F_{x_0} + \Delta y_i F_{y_0} + O\left( \Delta x_i, \Delta y_i \right)^2, \; \; i  = 1, \hdots, n
\label{taylor-expansion-F}
\end{equation}
where $\Delta x_i = x_i - x_0$, $\Delta y_i = y_i - y_0$, $\Delta F_i = F_i - F_0$ and $n$ represents the number of neighbours of the point $P_0$. For $n \geq 3$, eq. (\ref{taylor-expansion-F}) leads to an over-determined linear system for the unknowns $F_x$ and $F_y$, which can be solved using the least squares principle. The first-order accurate least squares approximations to $F_x$ and $F_y$ at the point $P_0$ are then given by
\begin{equation}
\begin{bmatrix}
F^1_x \\ F^1_y \end{bmatrix}
 = 
{\begin{bmatrix}
\sum \Delta x_i^2 &  \sum \Delta x_i \Delta y_i  \\
\sum \Delta x_i \Delta y_i & \sum \Delta y_i^2 
\end{bmatrix}}^{-1} 
\begin{bmatrix}
\sum \Delta x_i \Delta F_i \\
\sum \Delta y_i \Delta F_i \\
\end{bmatrix}
\label{ls-formulae-first-order}
\end{equation}
In the above formulae, the superscript $1$ on $F_x$ and $F_y$ denotes the first-order accuracy. Taking $ \boldsymbol{\Psi} $ - moments of eq. (\ref{kfvs-3d-be}) along with the formulae in eq. (\ref{ls-formulae-first-order}), we obtain the semi-discrete form of the first-order least squares kinetic upwind scheme for $2D$ Euler equations,
\begin{equation}
\frac{d \boldsymbol{U}}{d t} + \frac{\partial \boldsymbol{Gx}^+}{\partial x } + \frac{\partial \boldsymbol{Gx}^-}{\partial x } + \frac{\partial \boldsymbol{Gy}^+}{\partial y } + \frac{\partial \boldsymbol{Gy}^-}{\partial y }  = 0
\label{kfvs-ee}
\end{equation}
Here, $\boldsymbol{Gx}^{\pm}$ and $\boldsymbol{Gy}^{\pm}$ are the kinetic split fluxes \cite{lskum-ghosh-journal} along the $x$ and $y$ directions, respectively. The least squares formulae for the derivatives of $\boldsymbol{Gx}^{\pm}$ are given by 
\begin{equation}
\frac{\partial \boldsymbol{Gx}^{\pm}}{\partial x } 
 = \frac{1}{Det }
{\begin{vmatrix}
\sum \Delta x_i \Delta \boldsymbol{Gx}^{\pm}_i  &  \sum \Delta x_i \Delta y_i  \\
\sum \Delta y_i \Delta \boldsymbol{Gx}^{\pm}_i & \sum \Delta y_i^2 
\end{vmatrix}}, \; \; i \in N_x^{\pm}  \left(P_0 \right)
\label{ls-formulae-first-order-kfvs}
\end{equation}
Here, $Det$ is the determinant of the least squares matrix in eq. (\ref{ls-formulae-first-order}). 
Note that the above split flux derivatives are evaluated using the split stencils $N_x^{\pm}  \left(P_0 \right) $, defined by
\begin{equation}
N_x^{\pm} \left( P_0 \right)  = \left\lbrace P_i \mid P_i \in N \left( P_0 \right),  \Delta x_i = x_{i} - x_{0}  \lessgtr 0 \right\rbrace  
\end{equation}
Similarly, we can write the least squares formulae for the split flux derivatives of $\boldsymbol{Gy}^{\pm}$. Finally, the state-update formula for steady flow problems can be constructed by replacing the pseudo-time derivative in eq.~({\ref{kfvs-ee}}) with a suitable discrete approximation and local time stepping. In the present work, the solution is updated using a four-stage third-order Runge-Kutta (SSP-RK3) \cite{ssp-rk3} time marching algorithm.  
\subsection{Second-order accuracy using $\boldsymbol{q}$-variables}
One way of obtaining second-order accurate approximations to the spatial derivatives $F_x$ and $F_y$ is by considering the Taylor series expansion of $F$ up to quadratic terms,
\begin{equation}
\Delta F_i =   \Delta x_i F_{x_0} + \Delta y_i F_{y_0}  +  \frac{\Delta x_i^2}{2} F_{xx_0} + \Delta x_i \Delta y_i F_{xy_0} 
+  \frac{\Delta y_i^2}{2} F_{yy_0}  +  O \left( \Delta x_i, \Delta y_i \right)^3, \; \; i  = 1, \hdots, n
\label{taylor-series-exp-F-second-order}
\end{equation}
For $ n \geq 6$, we get an over-determined linear system of the form
\begin{equation}
\begin{bmatrix}
\Delta x_1      & \cdots &  \frac{\Delta y_1^2 }{2} \\
\Delta x_2       & \cdots &  \frac{\Delta y_2^2 }{2} \\
\vdots & \cdots & \vdots \\
\vdots &  \cdots &  \vdots \\
\Delta x_n       & \cdots &  \frac{\Delta y_n^2 }{2}
\end{bmatrix}
\begin{bmatrix}
F_{x_0} \\
 F_{y_0} \\
  \vdots \\
   \vdots \\
    F_{yy_0}
\end{bmatrix}  =
\begin{bmatrix}
\Delta F_1 \\ \Delta F_2 \\ \vdots \\ \vdots \\ \Delta F_n
\end{bmatrix}
\end{equation}
If we denote the coefficient matrix as $A$, the unknown vector as $\boldsymbol{dF}$, and the right-hand side vector as $\boldsymbol{ \Delta F}$, then the solution of the linear system using least squares is given by
\begin{equation}
 \boldsymbol{dF} = \left( A^T A \right)^{-1} \left( A^T \boldsymbol{ \Delta F} \right)
\end{equation}
The first two components of the vector $  \boldsymbol{dF} $ give the desired second-order approximations to $F_x$ and $F_y$. The formulae for $F_x$ and $F_y$ involve the inverse of a $5 \times 5$ least squares matrix $A^T A$. Central to the success of this formulation is that the least squares matrix should be well-conditioned. In the case of first-order approximation, the $2 \times 2 $ least squares matrix corresponding to a point becomes singular if and only if the points in its stencil lie on a straight line. On the other hand, the least squares matrix in the second-order formulae can become singular if the alignment of the stencil is such that at least two rows of the matrix $A^T A$ are linearly dependent. Furthermore, it lacks robustness as the least squares matrix corresponding to the boundary points can be poorly conditioned, which results in loss of accuracy. \\ \\
An efficient way of obtaining second-order accurate approximations to the spatial derivatives in eq. (\ref{kfvs-ee}) is by employing the defect correction method \cite{lskum-ghosh-journal}. An advantage of the defect correction procedure is that the dimension of the least squares matrix remains the same as in the first-order scheme. To derive the desired formulae, consider the Taylor expansion of $F$ up to quadratic terms, 
\begin{equation}
\begin{split}
\Delta F_i = & \Delta x_i F_{x_0} + \Delta y_i F_{y_0} 
+  \frac{\Delta x_i}{2} \left( \Delta x_i F_{xx_0} + \Delta y_i F_{xy_0}  \right)  
 +  \frac{\Delta y_i}{2} \left( \Delta x_i F_{xy_0} + \Delta y_i F_{yy_0}  \right)   \\
 + & O \left( \Delta x_i, \Delta y_i \right)^3, \; \; i  = 1, \hdots, n
\end{split}
\label{taylor-series-exp-F-second-order-rearranged}
\end{equation}
The basic idea of the defect correction procedure is to cancel the second-order derivative terms in the above equation by defining a modified $\Delta F_i$ so that the leading terms in the truncation errors of the formulae for $F_x$ and $F_y$ are of the order of $O \left( \Delta x_i, \Delta y_i  \right)^2$. Towards this objective, consider the Taylor series expansions of $F_x$ and $F_y$ up to linear terms
\begin{equation}
\begin{split}
\Delta F_{x_i} = &  \Delta x_i F_{xx_0} + \Delta y_i F_{xy_0} + O\left( \Delta x_i, \Delta y_i  \right)^2  \\
\Delta F_{y_i} = &  \Delta x_i F_{xy_0} + \Delta y_i F_{yy_0} + O\left( \Delta x_i, \Delta y_i \right)^2 
\end{split}
\end{equation}
where $\Delta F_{x_i} = F_{x_i} - F_{x_0} $ and $ \Delta F_{y_i} = F_{y_i} - F_{y_0} $. Using these expressions in eq. ~(\ref{taylor-series-exp-F-second-order-rearranged}), we obtain
\begin{equation}
\Delta F_i =  \Delta x_i F_{x_0} + \Delta y_i F_{y_0} +  \frac{1}{2} \left\lbrace \Delta x_i  \Delta F_{x_i}   +   \Delta y_i \Delta F_{y_i} \right\rbrace
 +  O \left( \Delta x_i, \Delta y_i \right)^3
\label{taylor-series-second-order-rearranged}
\end{equation}
We now introduce the modified perturbation in Maxwellians, $\Delta \widetilde{F}_i $ and define it as
\begin{equation}
\begin{split}
\Delta \widetilde{  F}_i = & \Delta F_i - \frac{1}{2} \left\lbrace \Delta x_i \Delta F_{x_i} +  \Delta y_i \Delta F_{y_i} \right\rbrace \\
 =  &\left\lbrace F_i - \frac{1}{2} \left( \Delta x_i F_{x_i} + \Delta y_i F_{y_i}  \right)  \right\rbrace
  -  \left\lbrace F_0 - \frac{1}{2} \left( \Delta x_i F_{x_0} + \Delta y_i F_{y_0} \right) \right\rbrace \\ 
= & \widetilde{  F}_i - \widetilde{  F}_0  
\end{split}
\label{modified-perturb-maxwellian}
\end{equation}
Using $\Delta \widetilde{ F}_i $, eq. (\ref{taylor-series-second-order-rearranged}) reduces to
\begin{equation}
\Delta \widetilde{ F}_i   =  \Delta x_i F_{x_0} +  \Delta y_i F_{y_0}  +  O \left( \Delta x_i, \Delta y_i \right)^3, \; \; i  = 1, \hdots, n
\label{taylor-series-exp-F-third-order-modified-final}
\end{equation}
Solving the above modified over-determined system using least squares, the second-order accurate approximations to $F_x$ and $F_y$ at the point $P_0$ are given by
\begin{equation}
\begin{bmatrix}
F_x^{2}\\ F_y^{2}  \end{bmatrix}
 = 
{\begin{bmatrix}
\sum \Delta x_i^2 &  \sum \Delta x_i \Delta y_i \\
\sum \Delta x_i \Delta y_i & \sum \Delta y_i^2  
\end{bmatrix}}^{-1} 
\begin{bmatrix}
\sum \Delta x_i  \Delta \widetilde{ F}_i  \\
\sum \Delta y_i \Delta \widetilde{ F}_i  \\
\end{bmatrix}
\label{ls-formulae-second-order}
\end{equation}
Note that superscript ${2} $ on $F_x$ and $F_y$ denotes second-order accuracy. 
The above formulae satisfy the test of $k$-exactness as they yield exact derivatives for polynomials of degree $\leq 2$. Furthermore, these formulae have the same structure as the first-order formulae in eq.~(\ref{ls-formulae-first-order}), except that the second-order approximations use modified Maxwellians. In contrast to first-order formulae that are explicit, the second-order approximations have implicit dependence. From eq. (\ref{ls-formulae-second-order}), the evaluation of $F_x$ and $F_y$ at the point $P_0$ requires the values of these derivatives at $P_0$ and its neighbours a priori.  \\ \\
Taking $\boldsymbol{\Psi}$-moments of the spatial terms in eq. (\ref{kfvs-3d-be}) along with the formulae in eq. (\ref{ls-formulae-second-order}), we get the second-order accurate discrete approximations for the kinetic split flux derivatives. For example, the expressions for the spatial derivatives of $\boldsymbol{Gx}^{\pm}$ are given by 
\begin{equation}
\frac{\partial \boldsymbol{Gx}^{\pm}}{\partial x } 
 = \frac{1}{Det}
{\begin{vmatrix}
\sum \Delta x_i \widetilde{\boldsymbol{Gx}}_i^{\pm}  &  \sum \Delta x_i \Delta y_i \\
\sum \Delta y_i \widetilde{\boldsymbol{Gx}}_i^{\pm} & \sum \Delta y_i^2 
\end{vmatrix}}
\label{ls-formulae-second-order-gxpm}
\end{equation}
The perturbations $\Delta \widetilde{\boldsymbol{Gx}}_i^{\pm}$ are defined by
\begin{equation}
\Delta \widetilde{\boldsymbol{Gx}}_i^{\pm}=  \Delta {\boldsymbol{Gx}}^{\pm}_i - \frac{1}{2} \left\lbrace   \Delta x_i \frac{\partial}{\partial x} \Delta \boldsymbol{Gx}_i^{\pm} +  \Delta y_i \frac{\partial}{\partial y} \Delta \boldsymbol{Gx}_i^{\pm} \right\rbrace  
\label{pert-mod-split-fluxes}
\end{equation}
A drawback of this formulation is that the second-order scheme thus obtained reduces to first-order at the boundaries as the stencils to compute the split flux derivatives may not have enough neighbours. Furthermore, $\Delta \widetilde{  F_i}$ is not the difference between two Maxwellians. Instead, it is the difference between two perturbed Maxwellians, $\widetilde{  F}_i$ and $\widetilde{  F}_0$. Unlike $F_i$ and $F_0$, the distribution functions $\widetilde{F}_i$ and $\widetilde{F}_0 $ may not be non-negative and thus need not be Maxwellians. \\ \\
In order to preserve positivity, instead of Maxwellians, we employ the $\boldsymbol{q}$-variables \cite{smd-nasa-report-entropy-variables,qlskum} in the defect correction procedure. The $\boldsymbol{q}$-variables in $2D$ are given by
\begin{equation}
\boldsymbol{q} = \begin{bmatrix}
\ln \rho + \frac{\ln \beta}{\gamma - 1} - \beta \left(u_1^2 + u_2^2 \right), \; 2 \beta u_1, \; 2 \beta u_2, \; -2\beta 
\end{bmatrix}    
\label{q_variables}
\end{equation}
Note that the transformations $F\longleftrightarrow \boldsymbol{q} $ and $\boldsymbol{U}  \longleftrightarrow \boldsymbol{q} $ are unique, and therefore the $\boldsymbol{q}$-variables can be used to represent the fluid flow at the macroscopic level. The second-order LSKUM based on $\boldsymbol{q}$-variables is then obtained by replacing $\Delta \widetilde{\boldsymbol{Gx}}_i^{\pm}$ in eq. (\ref{pert-mod-split-fluxes}) with $\Delta \boldsymbol{Gx}_i^{\pm} \left(\widetilde{\boldsymbol{q}}\right)$. The new perturbation in split fluxes is defined by
\begin{equation}
\Delta \boldsymbol{Gx}_i^{\pm} \left(\widetilde{\boldsymbol{q}}\right) =  \boldsymbol{Gx}^{\pm} \left(\widetilde{\boldsymbol{q}}_i\right) - \boldsymbol{Gx}^{\pm} \left(\widetilde{\boldsymbol{q}}_0\right) 
\end{equation}
 Here, $\widetilde{\boldsymbol{q}}_i$ and $\widetilde{\boldsymbol{q}}_0$ are the modified $\boldsymbol{q}$-variables, given by
\begin{equation}
\begin{split}
\widetilde{\boldsymbol{q}}_i & = \boldsymbol{q}_i - \frac{1}{2} \left( \Delta x_i {\boldsymbol{q}_x}_i +  \Delta y_i {\boldsymbol{q}_y}_i  \right)  \\
\widetilde{\boldsymbol{q}}_0 & = \boldsymbol{q}_0 - \frac{1}{2} \left( \Delta x_i {\boldsymbol{q}_x}_0 +  \Delta y_i {\boldsymbol{q}_y}_0   \right)  \\
\end{split}
\label{q-tilde-variables}
\end{equation}
The necessary condition for obtaining second-order accurate split flux derivatives is that the $\boldsymbol{q}$-derivatives in eq.~(\ref{q-tilde-variables}) should be second-order. Note that the $\boldsymbol{q}$-derivatives are approximated using least squares formulae with a full stencil as, 
\begin{equation}
    \begin{bmatrix}
    \boldsymbol{q}_x \\ \boldsymbol{q}_y
    \end{bmatrix} = {\begin{bmatrix}
\sum \Delta x_i^2 &  \sum \Delta x_i \Delta y_i \\
\sum \Delta x_i \Delta y_i & \sum \Delta y_i^2  
\end{bmatrix}}^{-1} 
\begin{bmatrix}
\sum \Delta x_i  \Delta \widetilde{ \boldsymbol{q}}_i  \\
\sum \Delta y_i \Delta \widetilde{\boldsymbol{q}}_i  \\
\end{bmatrix}
\label{ls-formulae-q-derivatives}
\end{equation}
The above formulae for $\boldsymbol{q}$-derivatives are implicit  and need to be solved iteratively. These sub-iterations are called inner iterations. In the present work, we perform numerical simulations with three inner iterations. \\ \\
%
An advantage of $\boldsymbol{q}$-variables is that higher-order accuracy can be achieved even at boundary points as the defect-correction procedure can be combined with the kinetic wall \cite{kfvs} and kinetic outer boundary \cite{kobc} conditions. Furthermore, the distribution functions $F\left(\widetilde{\boldsymbol{q}}_i \right) $ and $F\left(\widetilde{\boldsymbol{q}}_0 \right) $ corresponding to $\widetilde{\boldsymbol{q}}_i$ and $\widetilde{\boldsymbol{q}}_0$ are always Maxwellians and therefore preserves the positivity of numerical solution.  
\section{GPU accelerated meshfree q-LSKUM solver}
\label{gpu-solver}
In this section, we present the development of a GPU accelerated meshfree solver based on {\tt q-LSKUM}. We begin with a brief description of the steps required to compute the flow solution using a serial code. \\ \\
Algorithm {\ref{algo-serial}} presents a general structure of the serial meshfree {\tt q-LSKUM} solver for steady-state flows. The solver consists of a fixed point iterative scheme, where each iteration evaluates the local time step, four stages of the Runge-Kutta scheme, and the $L_2$ norm of the residue. The subroutine {\tt q\_variables()} evaluates the $\boldsymbol{q}$-variables defined in eq. (\ref{q_variables}) while {\tt q\_derivatives()} computes the second-order accurate approximations of $\boldsymbol{q}_x$ and $\boldsymbol{q}_y$ using the formulae in eq. (\ref{ls-formulae-q-derivatives}). The most time consuming routine is the {\tt flux\_residual()}, which performs the least squares approximation of the kinetic split flux derivatives in eq. (\ref{kfvs-ee}). {\tt state\_update(rk)} updates the flow solution at each Runge-Kutta step. All the input and output operations are performed in {\tt preprocessor()} and {\tt postprocessor()}, respectively. The parameter $N$ represents the number of pseudo-time iterations required to achieve a desired convergence in the flow solution. \\ \\
\begin{algorithm}[t]
    \DontPrintSemicolon
    \SetAlgoLined
    \vspace{1mm}
    \SetKwFunction{FMain}{q-LSKUM}
    \SetKwProg{Fn}{subroutine}{}{}
        \Fn{\FMain}
    {
                \vspace{1mm}
                call {\tt{ preprocessor() }}\\
                \vspace{1mm}
        \For{$n \leftarrow 1$ \KwTo $n \leq N$}
        {
            \vspace{1mm}
                     call {\tt{ timestep() }}\\
            \vspace{1mm}
            \For{$rk \leftarrow 1$ \KwTo $4 $}
            {
                \vspace{1mm}
                         call {\tt  q\_variables() } \\
                \vspace{1mm}
                         call    {\tt  q\_derivatives() } \\
                \vspace{1mm}
                         call {\tt  flux\_residual() } \\
                \vspace{1mm}
                         call  {\tt state\_update(rk) } \\
                \vspace{1mm}
            }
                        \vspace{1mm}
                        call {\tt  residue() } \\
                        \vspace{1mm}
        }
                \vspace{1mm}
                call {\tt{ postprocessor() }}\\
                \vspace{1mm}
    }
    \vspace{1mm}
    \textbf{end subroutine}
    \vspace{1mm}
    \caption{Serial meshfree solver based on q-LSKUM}
    \label{algo-serial}
\end{algorithm}
%
%
Algorithm \ref{algo-gpu} presents the structure of a GPU accelerated {\tt q-LSKUM} solver written in CUDA. The GPU solver mainly consists of the following sequence of operations: transfer the input data structure from host to device, performing fixed-point iterations on the device, and finally transfer the converged flow solution from device to host. In the current implementation, for each significant subroutine in the serial code, equivalent global kernels are constructed in the GPU code. %
\begin{algorithm}[htbp]
\SetAlgoLined
\vspace{2mm}
\caption{GPU accelerated meshfree solver based on q-LSKUM}
\vspace{2mm}
\SetKwFunction{FMain}{q-LSKUM}
\SetKwProg{Fn}{subroutine}{:}{}
    \Fn{\FMain}{
\vspace{1mm}    
                call {\tt{ preprocessor() }}\\    
\vspace{1mm}
    \textbf{cudaHostToDevice}(CPU\_data, GPU\_data)\\
\vspace{1mm}
\For{$n \leftarrow 1$ \KwTo $n \leq N$}{
\vspace{1mm}
         \textbf{Global kernel}  $\lll$  grid, block $\ggg$ {\tt timestep()} \\
\vspace{1mm}
\For{$rk \leftarrow 1$ \KwTo $4 $}{
\vspace{1mm}
         \textbf{Global kernel}  $\lll$  grid, block $\ggg$  {\tt q\_variables()} \\
\vspace{0.5mm}
         \textbf{Global kernel}  $\lll$  grid, block $\ggg$  { \tt q\_derivatives()} \\
\vspace{0.5mm}
         \textbf{Global kernel}  $\lll$  grid, block $\ggg$  { \tt flux\_residual()} \\
\vspace{0.5mm}
         \textbf{Global kernel}  $\lll$  grid, block $\ggg$  { \tt state\_update(rk)} \\
}
}
\vspace{1mm}
    \textbf{cudaDeviceToHost}(GPU\_data, CPU\_data)\\
\vspace{1mm}
call {\tt{ postprocessor() }}\\    
    }
\vspace{1mm}
\textbf{end subroutine}
     \label{algo-gpu}
\end{algorithm}
\section{Performance analysis of naive GPU solvers}
\label{sec-naive-gpu-codes-performance}
This section presents the numerical results to assess the performance of naive GPU solvers written in {\tt Fortran}, {\tt C++}, { \tt Python}, and {\tt Julia}. The test case under investigation is the inviscid fluid flow simulation around the NACA 0012 airfoil at Mach number, $M = 0.63$, and angle of attack, $AoA = 2^o$ . For the benchmarks, numerical simulations are performed on seven levels of point distributions. The coarsest distribution consists of $625,000$ points, while the finest distribution consists of $40$ million points.  \\ \\ 
Table {\ref{workstation-config}} shows the hardware configuration, while Table {\ref{gpu-code-specs}} shows the language specifications, compilers, and flags used to execute serial and GPU computations. The {\tt Python} GPU code uses Numba $0.55.0$ \cite{numba} and NumPy $1.20.1$ \cite{numpy}, while {\tt Julia} GPU code uses CUDA.jl $2.4.1$ library \cite{CUDAnative.jl-2018}. All the computations are performed with double precision using CUDA $11.2.2$. Appendix \ref{environment} presents the run-time environment and hardware specifications used to execute the serial and GPU codes. \\ 
\begin{table}[H]
\centering
    \begin{tabular}{lcc}
    \toprule
      & CPU   & GPU   \\
        \midrule
Model & AMD EPYC\textsuperscript{TM} $ 7542 $   & Nvidia Tesla $V100$ PCIe  \\ [0.3em]
Cores & $64$ $\left( 2 \times 32 \right)$ & $5120$ \\ [0.3em]
Core Frequency & $2.20$ GHz & $1.230$ GHz \\ [0.3em]
Global Memory & $256$ GiB & $32$ GiB \\ [0.3em]
$L2$ Cache & $16$ MiB & $6$ MiB \\ [0.1em]
        %
        \bottomrule
    \end{tabular}
    \caption{Hardware configuration used to perform numerical simulations. }
        \label{workstation-config}
\end{table}
\begin{table}[H]
\centering
    \begin{tabular}{lcccc}
    \toprule
    Language & Version & Compiler & Version & Flags \\
        \midrule
{\tt Fortran} & Fortran $90$ & nvfortran & $21.2$ & -O3 \\ [0.3em]
{\tt C++} & C++ $20$ & nvcc & $21.2$ & -O3 -mcmodel=large\\ [0.3em]
{\tt Python} & Python $3.9.1$ & Numba & $0.55.0$ & -O3 \\ [0.3em]
{\tt Julia} & Julia $1.5.3$ & CUDA.jl & $2.4.1$ & -O3 --check-bounds=no \\ [0.1em]
        \bottomrule
    \end{tabular}
    \caption{List of language and compiler specifications used to execute the codes. }
        \label{gpu-code-specs}
\end{table}
\subsection{RDP comparison of GPU Solvers}
\begin{figure}[H]
\centering
\includegraphics[scale=0.43,trim={5mm 0 10mm 0},angle=0,clip]{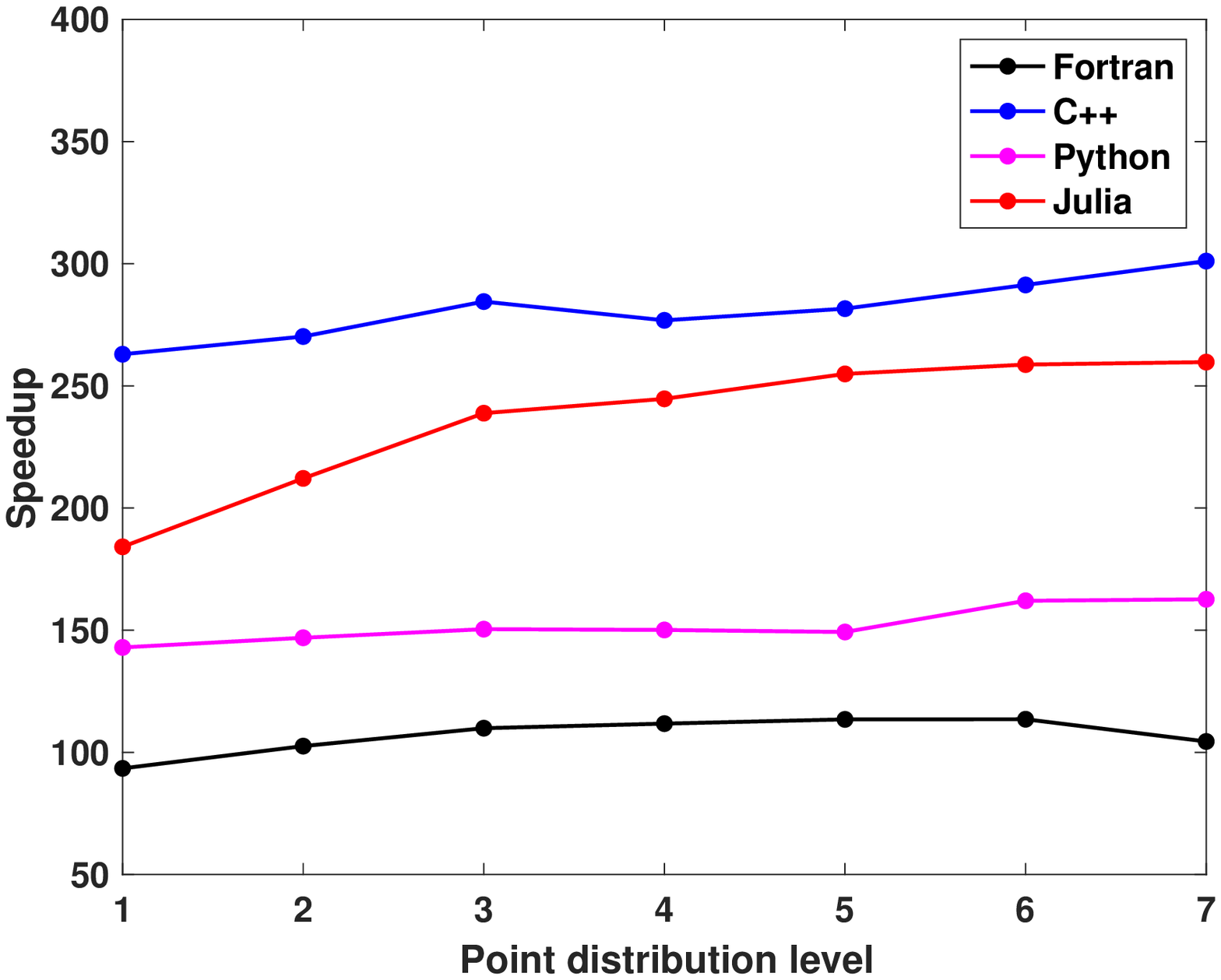} 
\includegraphics[scale=0.43,trim={5mm 0 10mm 0},angle=0,clip]{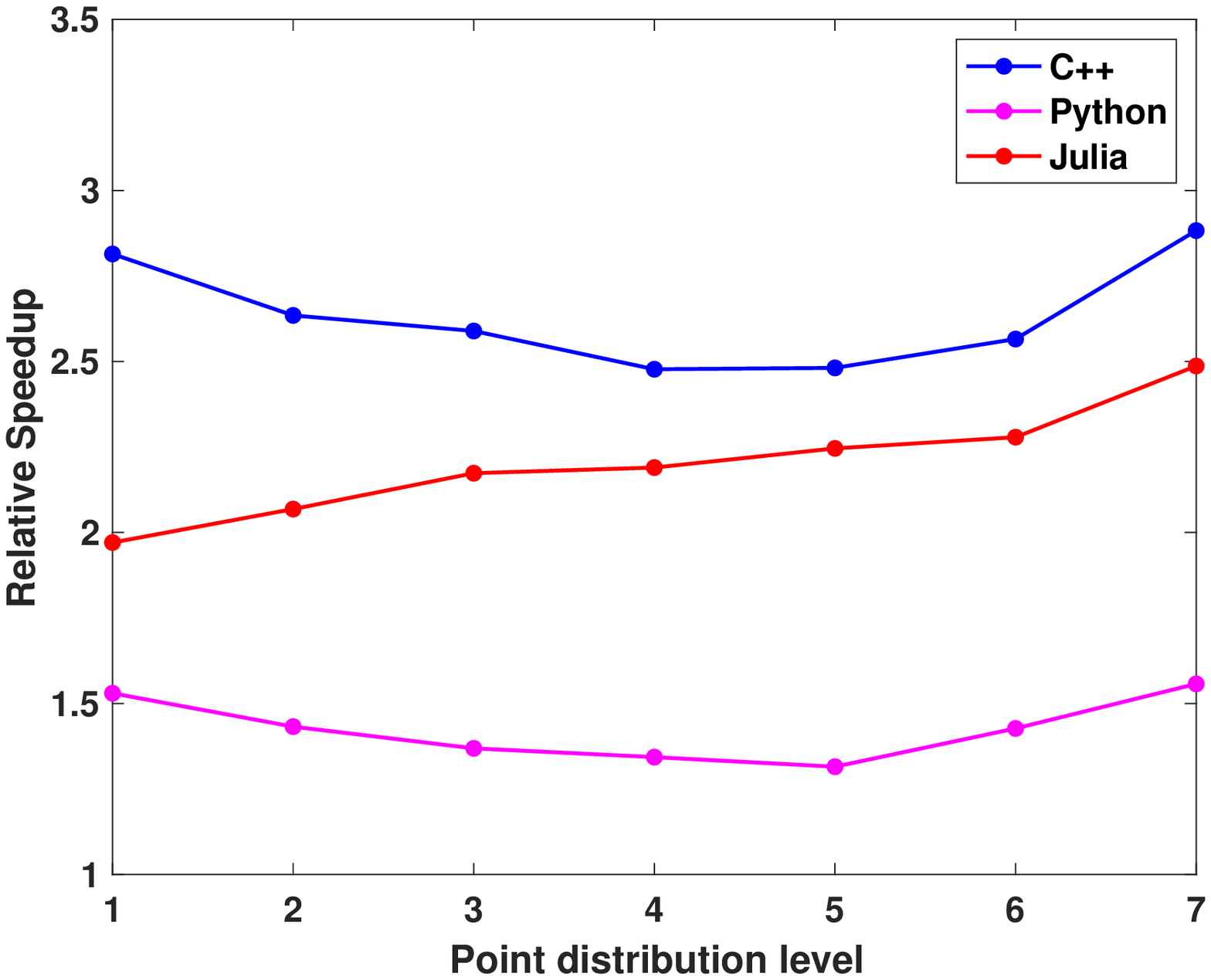} \\
(a)     \hspace{75mm}    (b) 
 \caption{ (a) Speedup achieved by the GPU codes. (b) Relative speedup of {\tt C++}, {\tt Python}, and {\tt Julia} GPU codes with respect to the {\tt Fortran} GPU code. }
     \label{speedup-naive-codes}
  \end{figure}

To measure the performance of the GPU codes, we adopt a cost metric called the Rate of Data Processing (RDP). The RDP of a meshfree code can be defined as the total wall clock time in seconds per iteration per point. Note that lower the value of RDP implies better the performance. Table {\ref{rdp-single-node}} shows a comparison of the RDP values for all the GPU codes. In the present work, the RDP values are measured by specifying the number of pseudo-time iterations in the GPU solvers to $1000$. For {\tt Fortran}, {\tt Python} and {\tt Julia} GPU codes, the optimal number of threads per block on all levels of point distribution is observed to be $64$. For {\tt C++}, the optimal number of threads per block is $128$. \\ \\
The tabulated values clearly show that the GPU solver based on {\tt C++} results in lowest RDP values on all levels of point distribution and thus exhibits superior performance. On the other hand, with the highest RDP values, the {\tt Fortran} code is computationally more expensive. As far as the {\tt Julia} code is concerned, its performance is better than {\tt Python} and closer to {\tt C++}. \\ \\
To assess the overall performance of the GPU meshfree solvers, we define another metric called speedup. The speedup of a GPU code is defined as the ratio of the RDP of the optimised serial {\tt C++} code to the RDP of the GPU code. Figure (\ref{speedup-naive-codes}a) shows the speedup achieved by the GPU codes, while (\ref{speedup-naive-codes}b) shows the relative speedup of {\tt C++}, {\tt Python} and {\tt Julia} GPU codes with respect to the {\tt Fortran} GPU code. From this figure, we observe that the {\tt C++} code is around $2.5$ times faster than {\tt Fortran}, while {\tt Julia} and {\tt Python} are respectively $2$ and $1.5$ times faster than {\tt Fortran}. 
\begin{table}[h]
\centering
\begin{tabular}{cccccc}
\toprule
Level & No. of points & Fortran & C++ & Python & Julia \\ [0.2em]
\midrule
 \multicolumn{6}{c}{RDP $\times$ $10^{-8}$ (Lower is better)}  \\ [0.2em]
\midrule
$1$ & $0.625$M  & $14.4090 $  & $5.1200 $ & $9.4183  $   & $7.3120 $  \\ [0.3em]
$2$ & $1.25$M   & $12.8570 $ & $ 4.8800 $ & $8.9765 $   & $ 6.2160 $  \\ [0.3em]
$3$ & $2.5$M   & $11.9100 $  & $4.6000 $  & $8.7008  $  & $ 5.4800 $  \\ [0.3em]
$4$ &  $5$M  & $ 11.5620 $ & $4.6673 $ & $ 8.6080 $ & $ 5.2800 $  \\  [0.3em]
$5$ & $10$M   & $11.3640 $ & $4.5800 $ & $ 8.6409  $ & $ 5.0600 $  \\ [0.3em]
$6$ & $20$M   & $11.3130 $ & $4.4096 $ & $ 7.9278 $ & $ 4.9650 $  \\ [0.3em]
$7$ & $40$M   & $12.2720 $ & $4.2573$ & $ 7.8805 $ & $ 4.9350 $  \\
\bottomrule
\end{tabular}
\caption{ Comparison of the RDP values based on naive GPU codes. }
\label{rdp-single-node}
\end{table}
\subsection{Run-time analysis of kernels}
To analyse the performance of the GPU accelerated meshfree solvers, it is imperative to investigate the kernels employed in the solvers. Towards this objective, NVIDIA \textsf{Nsight Compute} \cite{nvidia-documentation} is used to profile the GPU codes. Table {\ref{naive-gpu-codes-run-time-kernels}} shows the relative run-time incurred by the kernels on coarse, medium, and finest point distributions. Here, the relative run-time of a kernel is defined as the ratio of the kernel execution time to the overall time taken for the complete simulation.  \\ \\
This table shows that a very significant amount of run-time is taken by the {\tt flux\_residual} kernel, followed by {\tt q\_derivatives}. Note that the run-time of  {\tt q\_derivatives} kernel depends on the number of inner iterations. More the number of inner iterations, higher the time spent in its execution. For the kernels {\tt q\_variables} and {\tt state\_update}, the run-times are less than $2\%$ of the total execution time. For {\tt timestep} and {\tt host $\leftrightarrow$ device} operations, the run-times are found to be negligible and therefore not presented. \\ \\
%
 \\ \\
%
%
    \begin{table}[t]
        \centering
        \begin{center}
        \begin{tabular}{lccccc}
            \toprule
No.of points & Code    & q\_variables   & q\_derivatives & flux\_residual & state\_update \\[0.2em]
        \midrule
 & {\tt Fortran} & $0.50\%$ & $25.73\%$ & $ 72.67\%$ & $ 0.82 \%$  \\ [0.2em]
$0.625$M & {\tt C++} & $ 0.77 \%$ & $ 44.70\%$ & $ 50.51 \%$ & $1.87 \%$ \\ [0.2em]        
Coarse & {\tt Python} & $0.67\%$ & $37.48\%$ & $59.73\%$ & $1.47\%$ \\ [0.2em]        
& {\tt Julia} & $1.24\%$ & $24.52 \%$ & $71.71\%$ & $ 1.89 \%$  \\ [0.2em]        
        \cmidrule(lr){1-6}
 & {\tt Fortran} & $0.42\%$ & $25.60 \%$ & $ 72.95 \%$ & $ 0.74 \%$  \\ [0.2em]
$5$M & {\tt C++} & $ 0.80\%$ & $ 47.34\%$ & $47.68\%$ & $ 1.84\%$ \\ [0.2em]        
Medium & {\tt Python} & $0.60\%$ & $38.43\%$ & $59.10\%$ & $1.38 \%$ \\ [0.2em]        
& {\tt Julia} & $1.37 \%$ & $24.40 \%$ & $71.77\%$ & $ 1.85 \%$  \\ [0.2em]        
        \cmidrule(lr){1-6}
 & {\tt Fortran} & $0.41\%$ & $25.38\%$ & $ 73.21 \%$ & $ 0.74 \%$  \\ [0.2em]
$40$M & {\tt C++} & $0.81\%$ & $ 42.27\%$ & $ 52.94\%$ & $1.85\%$ \\ [0.2em]        
Fine & {\tt Python} & $0.58\%$ & $38.19\%$ & $59.40 \%$ & $ 1.35 \%$ \\ [0.2em]        
& {\tt Julia} & $1.32 \%$ & $24.12\%$ & $72.11 \%$ & $ 1.85\%$  \\ [0.2em]        
        \bottomrule
        \end{tabular}
        \end{center}
       \caption{{ Run-time analysis of the kernels on the finest point distribution. }}
        \label{naive-gpu-codes-run-time-kernels}
    \end{table}
\subsection{Performance metrics of the kernel {\tt flux\_residual}}
\begin{figure}[t]
\centering
\includegraphics[scale=0.5,angle=0,clip]{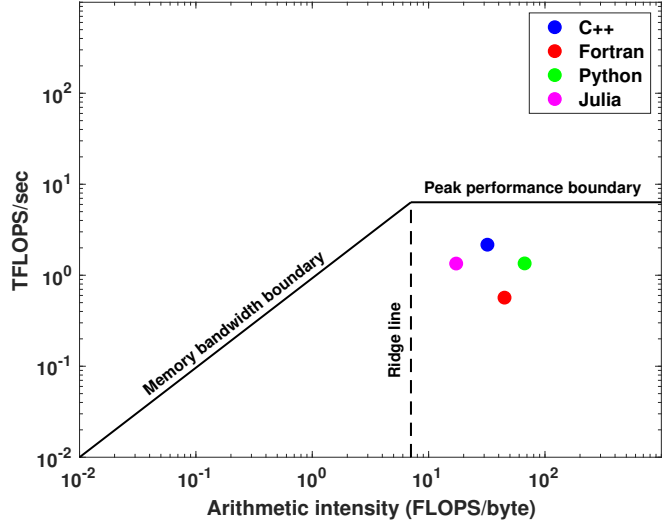} 
 \caption{ Roofline analysis of the {\tt flux\_residual} kernel. }
     \label{roofline-analysis-naive-codes}
  \end{figure}

To understand the varied run-times of the GPU codes in executing the {\tt flux\_residual} kernel, we investigate the kernel's utilisation of streaming multiprocessor (SM) and memory and achieved occupancy \cite{nvidia-documentation}. Table \ref{naive-performance-metrics} shows a comparison of these metrics on coarse, medium, and finest point distributions. We can observe that the {\tt C++} code has the highest utilisation of available SM resources, followed by {\tt Python} and {\tt Julia} codes. On the other hand, the {\tt Fortran} code has the poorest utilisation. Higher SM utilisation indicates an efficient usage of CUDA streaming multiprocessors, while lower values imply that the GPU resources are underutilised. In the present work, poor SM utilization limited the performance of the Fortran code as more time is spent in executing the flux-residual kernel. This resulted in higher RDP values for the {\tt Fortran} code. \\ \\ 
Table \ref{naive-performance-metrics} also presents the overall memory utilisation of the GPU codes. This metric shows the total usage of device memory. Furthermore, it also indicates the memory throughput currently being utilised by the kernel. Memory utilisation can become a bottleneck on the performance of a kernel if it reaches its theoretical limit \cite{nvidia-documentation}. However, low memory utilisation does not imply that the kernel optimally utilises it. The tabulated values show that the memory utilisation of the GPU codes is well within the acceptable limits.  \\ \\
To understand the poor utilisation of SM resources, we investigative the achieved occupancy of the {\tt flux\_residual} kernel. The achieved occupancy is the ratio of the number of active warps per SM to the maximum number of theoretical warps per SM. A code with high occupancy allows the SM to execute more active warps, thus increasing the overall SM utilisation. Low occupancy limits the number of active warps eligible for execution, leading to poor parallelism and latency. 
In the present work, all the GPU codes exhibited low occupancy for the {\tt flux\_residual} kernel. Table \ref{naive-performance-metrics} also compares register usage, one of the metrics that determine the number of active warps. In general, the higher the register usage, the lower the number of active warps. With the highest register usage, the {\tt Fortran} code has the lowest occupancy. \\ \\
The tabulated values have shown that the utilisation of SM and memory and achieved occupancy of the {\tt Python} code is higher than the {\tt Julia} code. However, the RDP values of {\tt Python} are much higher than {\tt Julia}. \\ \\
    \begin{table}[t]
        \centering
        \begin{center}
        \begin{tabular}{cccccc}
            \toprule
No.of points & Code    & SM & Memory   & Achieved  & Register usage\\
&  & utilisation &  utilisation  & occupancy & per thread\\[0.2em]
  \cmidrule(lr){3-5}
  & & \multicolumn{3}{c}{shown in percentage} & \\
        \midrule
 & {\tt Fortran} & $11.56 $ & $ 21.27 $  & $ 3.08$ & $220$\\ [0.2em]
$0.625$M & {\tt C++} & $43.16 $ & $ 10.41 $  & $11.76 $ & $184$\\ [0.2em]        
Coarse & {\tt Python} & $29.55 $ & $ 25.95 $  & $18.03$ & $128$\\ [0.2em]        
& {\tt Julia} & $ 26.23 $ & $ 18.28 $  & $ 16.54$ & $152$\\ [0.2em]        
        \cmidrule(lr){1-6}
 & {\tt Fortran} & $11.70 $ & $ 21.57$ &  $ 3.10 $ & $220$\\ [0.2em]
$5$M & {\tt C++} & $45.78 $ & $ 11.34$ & $12.03 $ & $184$\\ [0.2em]        
Medium & {\tt Python} & $ 30.05 $ & $26.35 $ & $ 18.29$ & $128$\\ [0.2em]        
& {\tt Julia} & $26.61$ & $18.15$ & $ 16.77$ & $152$\\ [0.2em]        
        \cmidrule(lr){1-6}
 & {\tt Fortran} & $11.68$ & $21.49$ &  $ 3.10 $ & $220$\\ [0.2em]
$40$M & {\tt C++} & $43.58$ & $ 9.15 $ &  $12.03$ & $184$\\ [0.2em]        
Fine & {\tt Python} & $30.31$ & $26.58$ & $ 18.33$ & $128$ \\ [0.2em]        
& {\tt Julia} & $ 27.10$ & $18.24$ & $ 16.76$ & $152$\\ [0.2em]        
        \bottomrule
        \end{tabular}
        \end{center}
\caption{{ A comparison of performance metrics on coarse, medium and finest point distributions. }}
        \label{naive-performance-metrics}
    \end{table}
To investigate this unexpected behaviour of the {\tt Python} code, we present the roofline analysis \cite{roofline} of the {\tt flux\_residual} kernel. A roofline model is a logarithmic plot that shows a kernel's arithmetic intensity with its maximum achievable performance. The arithmetic intensity is defined as the number of floating-point operations per byte of data movement. Figure \ref{roofline-analysis-naive-codes} shows the roofline analysis for all the GPU codes. Here, achieved performance is measured in trillions of floating-point operations per second. A code with performance closer to the peak boundary uses the GPU resources optimally. The {\tt C++} code, being closer to the roofline, yielded the best performance, while {\tt Fortran} is the farthest and resulted in poor performance. Although the achieved performance of {\tt Python} is the same as {\tt Julia}'s, its arithmetic intensity is much higher. Due to this, the RDP values of {\tt Python} are higher than {\tt Julia}. \\ \\
To investigate the difference in the utilisation of SM and memory, and the arithmetic intensity of {\tt Python} and {\tt Julia} codes, we analyse the scheduler and warp state statistics. 
Typically scheduler statistics consist of the metrics - {\it GPU maximum warps}, {\it active}, {\it eligible}, and {\it issued warps}. Here, {\it GPU maximum warps} is the maximum number of warps that can be issued per scheduler. For the NVIDIA V$100$ GPU card, the maximum warps is $16$. The warps for which resources such as registers and shared memory are allocated are known as {\it active warps}. {\it Eligible warps} are the subset of {\it active warps} that have not been stalled and are ready to issue their next instruction. From this set of {\it eligible warps}, the scheduler selects warps for which one or more instructions are executed. These warps are known as {\it issued warps}. Note that {\it active warps} is the sum of {\it eligible} and {\it stalled warps}. As far as the warp state statistics are concerned, it comprises several states for which warp stalls can occur. In the present work, the warp stalls due to {\it no instruction}, {\it wait}, and {\it long scoreboards} \cite{nvidia-documentation} are dominant. {\it No instruction} warp stall occurs when a warp is waiting to get selected to execute the next instruction. Furthermore, it can also happen due to instruction cache miss. In general, a cache miss occurs in kernels with many assembly instructions. A warp stalls due to {\it wait} if it is waiting for fixed latency execution dependencies such as fused multiply-add (FMA) or arithmetic-logic units (ALU). A {\it Long scoreboard} stall occurs when a warp waits for the requested data from {\tt L1TEX}, such as local or global memory units. If the memory access patterns are not optimal, then the waiting time for retrieving the data increases further. \\ \\
Table \ref{naive-performance-metrics-warps} shows the {\it scheduler statistics} on the finest point distribution. From this table we can observe that the {\tt C++} code has the lowest number of {\it active warps}. Although the number of {\it active warps} is more in {\tt Python} and {\tt Julia}, they are still much lesser than the {\it GPU maximum warps}. This is due to high register usage per thread in the {\tt flux\_residual} kernel. The tabulated values also show that the {\it eligible warps} are much less than the {\it active warps}, as most {\it active warps} are stalled. \\ \\
\begin{table}[t]
        \centering
        \begin{center}
        \begin{tabular}{cccccc}
            \toprule
            \multicolumn{1}{c}{No.of points} & \multicolumn{1}{c} {Code} & Active & Eligible & Issued & Eligible warps\\
\cmidrule(lr){3-5}
  & & \multicolumn{3}{c}{warps per scheduler  } & \multicolumn{1}{c}{in percentage}\\
        \midrule
        
$40$M & {\tt C++} & $1.93 $ & $ 0.24 $  & $0.21$ & $12.43 \%$ \\ [0.2em]        
Fine & {\tt Python} & $2.93 $ & $ 0.37 $  & $0.30$ & $12.62 \%$\\ [0.2em]        
& {\tt Julia} & $ 2.69 $ & $ 0.24 $  & $ 0.20$  & $8.92 \%$\\ [0.2em]        
        \bottomrule
        \end{tabular}
        \end{center}
\caption{{ A comparison of scheduler statistics on the finest level of point distribution. }}
        \label{naive-performance-metrics-warps}
\end{table}
We investigate the {\it warp state statistics} to understand the reason behind the low eligible warps in the {\tt flux\_residual} kernel. Table \ref{naive-performance-metrics-warp-stalls} shows a comparison of stall statistics measured in cycles. Note that the cycles spent by a warp in a stalled state define the latency between two consecutive instructions. These cycles also describe a warp's readiness or inability to issue the next instruction. The larger the cycles in the warp stall states, the more warp parallelism is required to hide latency. The tabulated values show that the overall stall in warp execution is maximum for {\tt Julia}. Due to this, {\tt Julia} has the lowest percentage of eligible warps. \\ \\
%
\begin{table}[t]
        \centering
        \begin{center}
        \begin{tabular}{ccccc}
            \toprule
            \multicolumn{1}{c}{No.of points} &
            \multicolumn{1}{c}{Code} & 
            \multicolumn{3}{c}{Stall in warp execution (in cycles) due to}
            \\
            \cmidrule(lr){3-5} 
&  &  no instruction &   wait  & long scoreboard \\ [0.2em]
        \midrule
$40$M & {\tt C++} & $2.96 $ & $ 3.12 $  & $0.87 $ \\ [0.2em]       
Fine & {\tt Python} & $4.94 $ & $ 2.14 $  & $0.66$ \\ [0.2em]       
& {\tt Julia} & $ 5.4 $ & $ 2.6 $  & $ 3.10$ \\ [0.2em]       
        \bottomrule
        \end{tabular}
        \end{center}
\caption{{ A comparison of warp state statistics  on the finest level of point distribution. }}
        \label{naive-performance-metrics-warp-stalls}
\end{table}
The scheduler and warp state statistics analysis did not reveal any conclusive evidence regarding the poor performance of {\tt Python} code over {\tt Julia}. To further analyse, we shift our focus towards the instructions executed inside the warps. In this regard, we investigate the global and shared memory access patterns of the warps and the pipe utilisation of the SM. \\ \\
Table \ref{naive-performance-metrics-global-load-store} shows a comparison of metrics related to global memory access. Here, global load corresponds to the load operations to retrieve the data from the global memory. In contrast, global store refers to the store operations to update the data in the global memory. A {\it sector} is an aligned $32$ byte-chunk of global memory. The metric, {\it sectors per request}, is the average ratio of sectors to the number of load or store operations by the warp. Note that the higher the {\it sectors per request}, the more cycles are spent processing the load or store operations. We observe that the {\tt Python} code has the highest number of {\it sectors per request} while {\tt Julia} has the lowest values. With the highest number of {\it sectors per request}, the {\tt Python} code suffers from poor memory access patterns. \\ \\
\begin{table}[t]
        \centering
        \begin{center}
        \begin{tabular}{ccccc}
            \toprule
            \multicolumn{1}{c}{Code} & 
            \multicolumn{2}{c}{Global Load} &
            \multicolumn{2}{c}{Global Store} \\ [0.2em]
            \cmidrule(lr){2-3} \cmidrule(lr){4-5} 
& Sectors & Sectors per request & Sectors & Sectors per request\\ [0.2em]
        \midrule
{\tt C++} & $3,789,109,860$ & $10.63$ & $43,749,721$ & $8.75$ \\ [0.2em]       
{\tt Python} & $14,637,012,265$ & $26.92$ & $159,999,732$ & $32.00$ \\ [0.2em]       
{\tt Julia} & $7,884,258,310$ & $7.41$ & $40,000,000$ & $8.00$ \\ [0.2em]       
        \bottomrule
        \end{tabular}
        \end{center}
\caption{ A comparison of global load and store metrics on the finest level of point distribution. }
        \label{naive-performance-metrics-global-load-store}
\end{table}
Table \ref{naive-performance-metrics-shared-bank-conflicts} shows a comparison of shared memory bank conflicts for {\tt C++}, {\tt Python}, and {\tt Julia} codes. A bank conflict occurs when multiple threads in a warp access the same memory bank. Due to this, the load or store operations are performed serially. The {\tt C++} code does not have any bank conflicts, while {\tt Julia} has bank conflicts due to load operations only. The {\tt Python} code has a significantly large number of bank conflicts and thus resulted in the poor performance of the {\tt flux\_residual} kernel.  \\ \\
\begin{table}[t]
        \centering
        \begin{center}
        \begin{tabular}{cccc}
            \toprule
            \multicolumn{1}{c}{No.of points} &
            \multicolumn{1}{c}{Code} & 
            \multicolumn{2}{c}{Shared memory bank conflicts due to} \\ [0.2em]
            \cmidrule(lr){3-4}
 & & load operations & store operations\\ [0.2em]
        \midrule
& {\tt C++} & $0$ & $0$ \\ [0.2em]       
$40$M & {\tt Python} & $3,824,672$ & $107,628,065$ \\ [0.2em]       
& {\tt Julia} & $4,413,868$ & $0$ \\ [0.2em]       
        \bottomrule
        \end{tabular}
        \end{center}
\caption{{ A comparison of shared memory bank conflicts due to load and store operations. }}
        \label{naive-performance-metrics-shared-bank-conflicts}
\end{table}
Table \ref{naive-performance-metrics-pipe-utilisation} shows the utilisation of dominant pipelines such as double-precision floating-point (FP64), Fused Multiply Add (FMA), Arithmetic Logic Unit (ALU), and Load Store Unit(LSU) for the {\tt flux\_residual} kernel. The FP64 unit is responsible for executing instructions such as {\tt DADD}, {\tt DMUL}, and {\tt DMAD}. A code with a high FP64 unit indicates more utilisation of 64-bit floating-point operations. The FMA unit handles instructions such as {\tt FADD}, {\tt FMUL}, {\tt FMAD}, etc. This unit is also responsible for integer multiplication operations such as {\tt IMUL}, {\tt IMAD}, and integer dot products. The ALU is responsible for the execution of logic instructions. The LSU pipeline issues load, store, atomic, and reduction instructions for global, local, and shared memory. The tabulated values show that {\tt Python} and {\tt Julia} codes have similar FP64 and LSU utilisation. However, the {\tt Python} code has excessive utilisation of FMA and ALU. This is due to the {\tt Numba} JIT compiler, which is not generating optimal SASS code for the {\tt flux\_residual} kernel. \\ \\
\begin{table}[t]
        \centering
        \begin{center}
        \begin{tabular}{ccccc}
            \toprule
            \multicolumn{1}{c}{Code} & 
            \multicolumn{1}{c}{Double-precision} &
            \multicolumn{1}{c}{Fused Multiply} &
            \multicolumn{1}{c}{Arithmetic Logic} &
            \multicolumn{1}{c}{Load Store} \\ [0.2em]
 & floating-point (FP64) & Add (FMA) & Unit (ALU) & Unit (LSU)\\ [0.2em]
        \midrule
{\tt C++} & $43.63$ & $6.58$ & $5.87$ & $1.78$ \\ [0.2em]       
{\tt Python} & $28.67$ & $14.28$ & $21.24$ & $8.05$ \\ [0.2em]
{\tt Julia} & $27.09$ & $9.41$ & $9.43$ & $7.97$ \\ [0.2em]       
        \bottomrule
        \end{tabular}
        \end{center}
\caption{ A comparison of pipe utilisation of the streaming multiprocessor (SM). }
        \label{naive-performance-metrics-pipe-utilisation}
\end{table}
To analyse the excessive utilisation of the FMA and ALU pipelines in {\tt Python}, Table \ref{naive-performance-metrics-instructions} compares the dominant instructions executed on the SM. We can observe that the {\tt Python} code has generated an excessive number of {\tt IMAD} and {\tt IADD3} operations that are not part of the meshfree solver. The additional instructions are generated due to CUDA thread indexing. This hampered the overall performance of the {\tt Python} code. \\ \\
In summary, the {\tt C++} code with better utilisation of SM yielded the lowest RDP values. The {\tt Fortran} code with very low occupancy resulted in the highest RDP. The {\tt Python} code has better utilisation of SM and memory and achieved occupancy compared to {\tt Julia}. However, it suffers from global memory coalescing, shared memory bank conflicts, and excessive utilisation of FMA and ALU pipelines. Due to this, the RDP values of the {\tt Python} code are significantly higher than the {\tt Julia} code. 
\begin{table}[t]
        \centering
        \begin{center}
        \begin{tabular}{ccccccc}
            \toprule
            \multicolumn{1}{c}{No.of points} &
            \multicolumn{1}{c}{Code} & 
            \multicolumn{1}{c}{DFMA} &
            \multicolumn{1}{c}{IMAD} &
            \multicolumn{1}{c}{DMUL} &
            \multicolumn{1}{c}{IADD3} &
            \multicolumn{1}{c}{DADD} \\ [0.2em]
        \midrule
\multicolumn{7}{c}{Instructions presented in Billions}  \\ [0.2em] 
        \midrule
&{\tt C++} & $6.1262  $ & $2.7451$ & $2.0509 $ & $0.9514 $ & $1.4174 $ \\ [0.2em]       
$40$M&{\tt Python} & $8.2769$ & $14.1171$ & $2.3879$ & $4.1338$ & $3.1966 $ \\ [0.2em]
&{\tt Julia} & $6.3009$ & $6.8711 $ & $2.2617 $ & $2.6878$ & $1.4201 $ \\ [0.2em]       
        \bottomrule
        \end{tabular}
        \end{center}
\caption{ A comparison of various instructions executed by an SM. }
        \label{naive-performance-metrics-instructions}
\end{table}
\section{Performance analysis of optimised GPU solvers}
\label{sec-opt-gpu-codes-performance}
\begin{figure}[H]
\centering
\includegraphics[scale=0.43,trim={5mm 0 10mm 0},angle=0,clip]{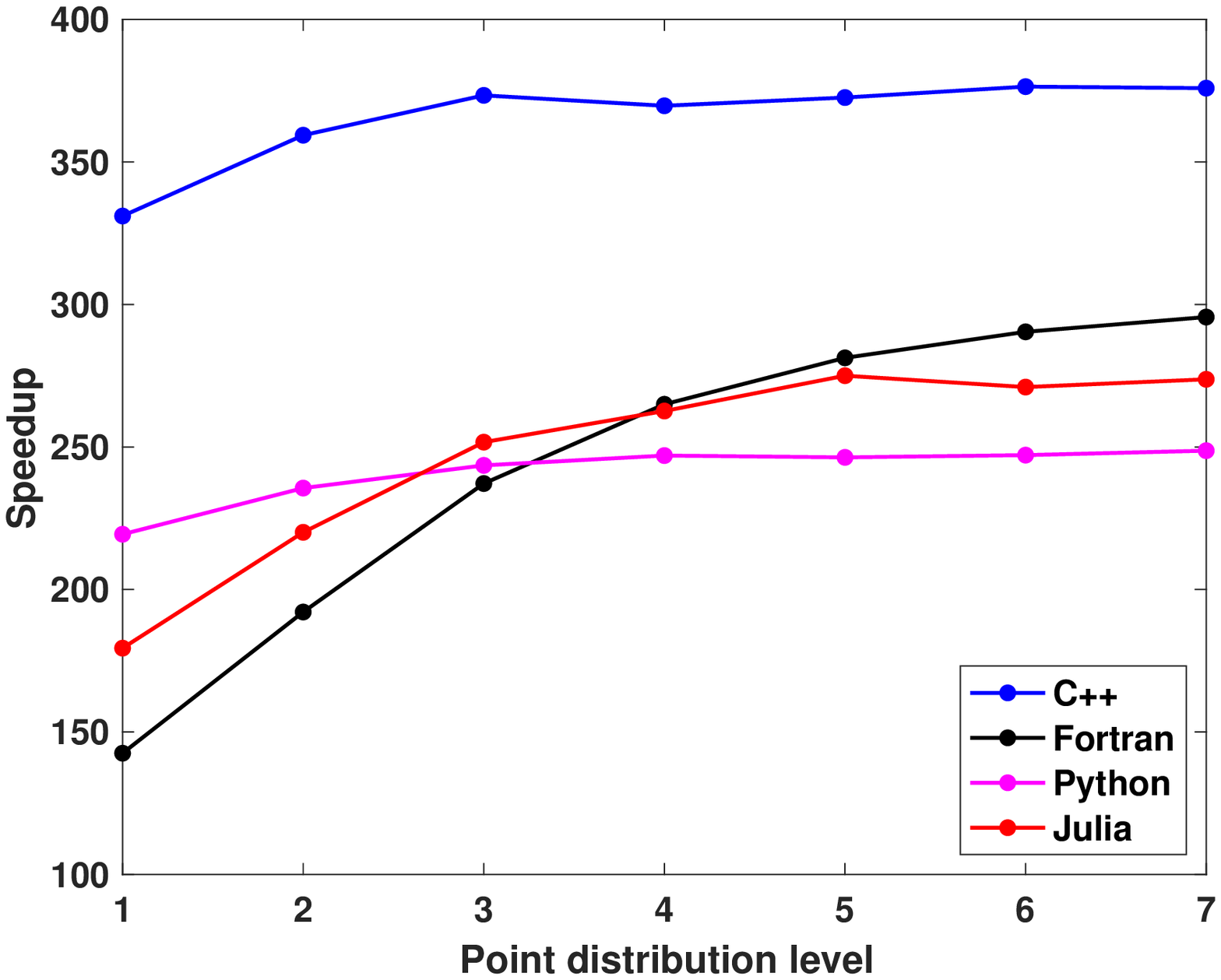} 
\includegraphics[scale=0.43,trim={5mm 0 10mm 0},angle=0,clip]{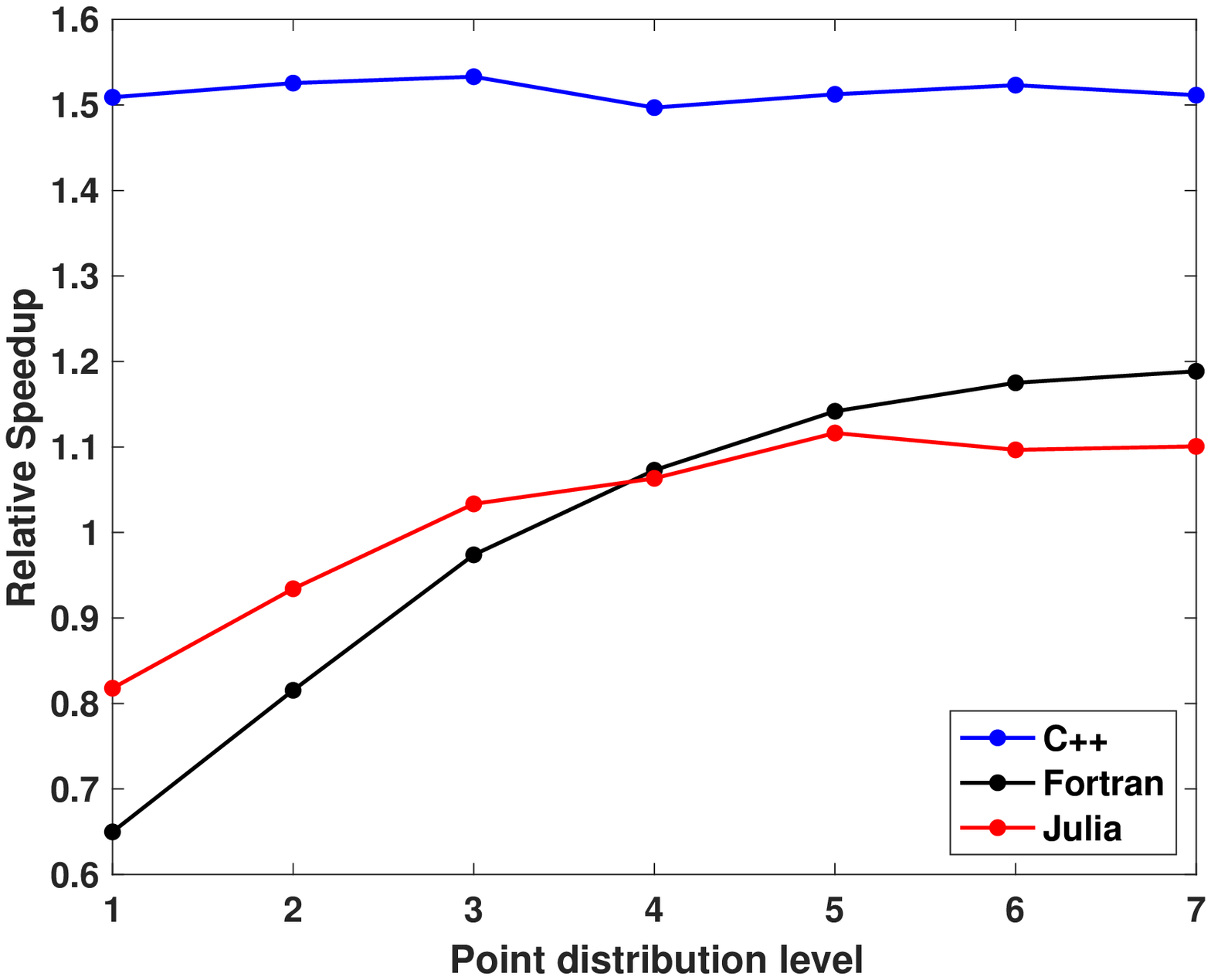} \\
(a)     \hspace{75mm}    (b) 
 \caption{ (a) Speedup achieved by the optimised GPU codes. (b) Relative speedup of optimised {\tt C++}, {\tt Fortran}, and {\tt Julia} GPU codes with respect to the optimised {\tt Python} GPU code. }
     \label{speedup-optimised-gpu-codes}
  \end{figure}

The analysis of several performance metrics has shown that there is scope for further improvement in the computational efficiency of the {\tt flux\_residual} kernel. Towards this objective, various optimisation techniques have been employed.  \\ \\
The profiler metrics have shown that the register usage of the {\tt flux\_residual} kernel is very high, which indicates that the size of the kernel is too large. To circumvent this problem, the {\tt flux\_residual} kernel is split into four smaller kernels that compute the spatial derivatives of the split fluxes $\boldsymbol{Gx}^+ $, $\boldsymbol{Gx}^- $, $\boldsymbol{Gy}^+ $ and $\boldsymbol{Gy}^- $, respectively. Note that these kernels are of similar size. In general, a smaller kernel consumes fewer registers compared to a larger kernel. Furthermore, kernels that are limited by registers will have an improved occupancy. Table \ref{optimsed-codes-global-sectors} shows a comparison of register usage per thread, achieved occupancy, and global sectors per request for the naive and optimised GPU codes. We present a range for metrics with both a lower and an upper bound for all the split flux kernels of the optimised codes. For the optimised codes, we present a range for metrics that has both a lower and an upper bound for all the split flux kernels. The tabulated values show a significant decrease in the register usage of the {\tt Fortran} code, followed by {\tt C++} and {\tt Julia}. In the case of {\tt Python}, the reduction is observed to be marginal. We also observe that the smaller kernels have more achieved occupancy compared to the {\tt flux\_residual} kernel. However, in the case of {\tt Python}, the occupancy did not improve as it is limited by the shared memory required per thread block. \\ \\ 
    \begin{table}[t]
        \centering
        \begin{center}
        \begin{tabular}{lccccc}
            \toprule
Number & Code    & Register usage & Achieved    & \multicolumn{2}{c}{Global sectors per request} \\
\cmidrule(lr){5-6}
of points &  & per thread &  occupancy  &  Load & Store\\[0.2em]
        \midrule
 & {\tt Fortran} - naive & $220$ & $ 3.10 $ & $24.34$ & $31.56$\\ [0.1em]
  &  {\tt Fortran} - optimised  &  $156$  &  $17.84-18.10$ & $17.86-18.25$ & $7.11$\\ [0.6em]
 & {\tt C++} - naive & $184$ &  $12.03$ & $10.63$ & $8.75$\\ [0.1em]        
$40$M &  {\tt C++} - optimised  & $154$ & $17.81-18.10$ & $10.19-10.31$ & $8.75$\\ [0.6em]        
 Fine & {\tt Python} - naive & $128$ & $18.33$ & $26.92$ & $32.00$\\ [0.1em]
 &  {\tt Python} - optimised  & $122$ & $17.87-18.16$ & $26.30-26.51$ & $32.00$\\ [0.6em]        
& {\tt Julia} - naive & $152$ & $ 16.76$ & $6.29$ & $4.37$\\ [0.1em]        
&  {\tt Julia} - optimised &  $128$ & $23.69-24.02$ & $6.26-6.31$ & $4.42$\\ [0.2em]
        \bottomrule
        \end{tabular}
        \end{center}
        \caption{{ A comparison of register usage, occupancy, and global sector per request of the naive and optimised GPU codes. }}
        \label{optimsed-codes-global-sectors}
    \end{table} 
To further enhance the computational efficiency of the kernels, the following language-specific optimisations are implemented. For the {\tt Fortran} code, instead of accessing and updating the arrays in an iterative loop, array slices are used. This improved the memory access patterns and global memory coalescing. Table \ref{optimsed-codes-global-sectors} clearly shows a reduction in {\tt Fortran}'s load and store operations, which increases the SM utilisation. However, this optimisation technique does not apply to the {\tt C++} code, as arrays are used instead of vectors. It is also not applicable for {\tt Julia} code, where values are accessed individually from a two-dimensional array. In the case of {\tt Python}, the current Numba compiler is unable to compile kernels that use array slices. 
%
For {\tt Fortran}, {\tt Python}, and {\tt Julia} codes, thread index and block dimensions are used to access values stored in the shared memory. This approach optimised the array indexing and allowed the threads to access the memory without any bank conflicts. However, in {\tt C++} code, shared memory is not used, and therefore, the above strategy is not applicable. Appendix \ref{appendix-indexing} presents an example code in {\tt Python} with naive and optimised versions of indexing for shared memory arrays. \\ \\
All the above optimisation techniques, except kernel splitting are implemented in other kernels wherever applicable. Table \ref{naive-gpu-codes-run-time-kernels} shows that, after {\tt flux\_residual}, {\tt q\_derivatives} is the most computationally intensive kernel. Splitting of this kernel is not feasible as  $\boldsymbol{q}$-derivatives in eq. (\ref{ls-formulae-q-derivatives}) are evaluated implicitly. Note that these optimisations may not yield a considerable reduction in the RDP values of smaller point distributions. However, on finest point distributions involving millions of points, these changes will significantly reduce the RDP values. \\ \\
%
Table \ref{optimised-codes-SM} shows a comparison of SM utilisation, performance in {\tt TFLOPS}, and arithmetic intensity. Compared to the {\tt flux\_residual} kernel, the split flux kernels have more SM utilisation and thus resulted in more {\tt TFLOPS}. For the split kernels based on {\tt Fortran}, {\tt C++}, and, {\tt Python} the arithmetic intensity is to the right of the ridgeline value of $7.05$. This implies that the kernels of these codes are compute bounded. On the other hand, the {\tt Julia} code is memory bounded as the arithmetic intensity of its split kernels lies to the left of the ridgeline. \\ \\
Table \ref{rdp-optimised-codes} shows a comparison of RDP values based on the optimised GPU codes. Note that for all the optimised codes, the optimal number of threads per block is $128$. We can observe that the optimisation has significantly enhanced the efficiency of the codes and thus resulted in smaller RDP values. The {\tt C++} code has the lowest RDP values on all levels of point distribution, followed by {\tt Fortran}. Although optimisation techniques have reduced the RDP values of the {\tt Python} code, they are till higher than the {\tt Julia} code. Figure (\ref{speedup-optimised-gpu-codes}a) shows the speedup achieved by the optimised codes, while (\ref{speedup-optimised-gpu-codes}b) shows the relative speedup of optimised {\tt C++}, {\tt Fortran}, and {\tt Julia} GPU codes with respect to the {\tt Python} GPU code. From this figure, on the finest point distribution, the {\tt C++} code is around $1.5$ times faster than the {\tt Python} code, while {\tt Fortran} and {\tt Julia} codes are faster than {\tt Python} by $1.2$ and $1.1$ times respectively. 
    \begin{table}[H]
        \centering
        \begin{center}
        \begin{tabular}{lcccc}
            \toprule
Number & Code & SM & Performance  & Arithmetic \\
of points & & utilisation & measured in {\tt TFLOPS}  & intensity \\[0.2em]
        \midrule
 & {\tt Fortran} - naive & $11.68$ & $0.5675$ & $44.89$ \\ [0.1em]
  &  {\tt Fortran} - optimised &  $47.85-48.68$ & $ 2.3547 - 2.4120$ & $10.71 - 10.90$  \\ [0.6em]
 & {\tt C++} - naive & $43.58$ & $ 2.1664$ & $32.00$ \\ [0.1em]        
$40$M &  {\tt C++} - optimised &  $56.41-58.30$ & $2.7947 - 2.8830 $  & $9.12 - 9.66$\\ [0.6em]        
 Fine & {\tt Python} - naive & $30.31$ & $1.3491$ & $66.84$ \\ [0.1em]
 &  {\tt Python} - optimised  & $54.29-55.36$ & $ 2.5794 - 2.6425$ & $18.20 - 18.30$\\ [0.6em]        
& {\tt Julia} - naive & $27.10$ & $ 1.3443$ & $17.25$\\ [0.1em]        
&  {\tt Julia} - optimised  & $34.19-34.42$ & $ 1.6862 - 1.6990$ & $4.93 - 7.93$ \\ [0.2em]
        \bottomrule
        \end{tabular}
        \end{center}
        \caption{ A comparison of SM utilisation, performance, and arithmetic intensity of the naive and optimised GPU codes. }
        \label{optimised-codes-SM}
    \end{table}
\begin{table}[H]
\centering
\begin{tabular}{cccccc}
\toprule
No. of points & Version & Fortran & C++ & Python & Julia \\ [0.2em]
\midrule
 \multicolumn{6}{c}{RDP $\times$ $10^{-8}$ (Lower is better)}  \\ [0.2em]
\midrule
$0.625$M  & naive & $14.4090 $  & $5.1200 $ & $9.4183  $   & $7.3120 $ \\ [0.2em]
 $0.625$M & optimised   & $9.4446$ & $4.0671 $ &  $ 6.1372 $ &  $ 7.5040 $ \\ [0.8em]
$5$M & naive   & $11.5620 $  & $4.6673$  & $8.6080 $  & $ 5.2800 $  \\ [0.2em]
 $5$M & optimised & $4.5856 $ & $3.4616$ & $ 5.2355 $ & $ 4.6900$  \\ [0.8em]
$40$M & naive   & $12.2720 $ & $4.2573$ & $ 7.8805 $ & $ 4.9350 $  \\[0.2em]
 $40$M & optimised  &  $4.3365$ & $3.4100 $ & $ 5.1540 $ & $ 4.6825 $  \\
\bottomrule
\end{tabular}
\caption{ A comparison of the RDP values based on naive and optimised GPU codes. }
\label{rdp-optimised-codes}
\end{table}
%
%
%
%
\section{Conclusions }
\label{sec-conclusions}
In this report we have presented an analysis of the performance of GPU accelerated meshfree solvers for compressible fluid flows in {\tt Fortran}, {\tt C++}, {\tt Python}, and {\tt Julia}. The meshfree solver was based on the least squares kinetic upwind method with entropy variables (q-LSKUM). The performance of the GPU codes was assessed on seven levels of point distribution ranging from $0.625$ million to $40$ million points. The performance of the solvers was measured by introducing a metric called the rate of data processing (RDP). Benchmark simulations have shown that the {\tt C++} GPU code resulted in the best performance with the smallest RDP values followed by {\tt Julia} and {\tt Python} codes. On the other hand, the {\tt Fortran} code was computationally more expensive with the highest RDP values. \\ \\
To investigate the differences in the RDP values of the GPU codes, the run-time analysis of the kernels was performed. The {\tt flux\_residual} kernel was observed to be dominant with the maximum time spent in its execution. This kernel was profiled using {\tt Nsight} to capture various performance metrics such as SM and memory utilisation, achieved occupancy, and registers per thread. The {\tt Fortran} code with high register usage resulted in low occupancy, which reduced its SM utilisation and thus resulted in high RDP values. The {\tt C++} code with the highest SM utilisation achieved the lowest RDP values. Although the utilisation of SM and memory and occupancy of the {\tt Python} code is higher than the {\tt Julia} code, its RDP values are much higher than the {\tt Julia} code. The roofline analysis of the {\tt flux\_residual} kernel was presented to investigate this behaviour. From this analysis, it was observed that the {\tt Python} and {\tt Julia} codes have almost the same achieved performance. However, the arithmetic intensity of {\tt Python} code was much higher than {\tt Julia}. Due to this, the RDP values of the {\tt Python} code were much higher than the {\tt Julia} code. \\ \\ %
The scheduler and warp state statistics were analyzed to explore the difference in the utilization of SM and memory and arithmetic intensity of the {\tt Python} and {\tt Julia} codes. However, it did not reveal any conclusive evidence regarding the poor performance of the {\tt Python} code. To investigate further, global load-store operations, shared memory bank conflicts, pipeline utilisation, and instructions executed by an SM were presented. For the {\tt Python} code, the global sectors per request, shared memory bank conflicts, FMA, and ALU pipelines were much higher than {\tt Julia}. Due to this, the RDP values of {\tt Python} were higher than {\tt Julia}.\\ \\
To further enhance the computational efficiency and thereby reducing the RDP of the GPU solvers, the {\tt flux\_residual} kernel was split into four equivalent smaller kernels. Kernel splitting has reduced the register pressure and improved the occupancy and overall SM utilisation. Few language-specific optimisations further enhanced the performance of the codes. \\ \\
Post optimisation, the {\tt Fortran} code was more efficient than {\tt Python} and {\tt Julia} codes. However, the {\tt C++} code is still the most efficient as the SASS code generated by its compiler is optimal compared to other codes. The optimised {\tt Python} code was computationally more expensive as the SASS code generated by the Numba compiler was not efficient. \\ \\ 
In the future, we plan to extend the GPU codes to compute three-dimensional flows of interest to aerospace engineering. Work is in progress in porting the codes to multi GPUs. 
%
%
\bibliographystyle{plain}
\bibliography{main}
\newpage
\section{Appendix}
\subsection{Example code for indexing shared memory arrays}
\label{appendix-indexing}
%
%
\begin{longlisting}
\begin{minted}{python}
@cuda.jit
def example_kernel(a):
    tx = cuda.threadIdx.x
    bx = cuda.blockIdx.x
    bw = cuda.blockDim.x # 128 threads per block

    temp = cuda.shared.array(shape = (128 * 4), dtype=numba.float64)

    for i in range(4):
        temp[cuda.threadIdx.x * 4 + i] = a[i]
\end{minted}
\caption{Original Code - Python}
\label{python-original-code}
\end{longlisting}
\begin{longlisting}
\begin{minted}{python}
@cuda.jit
def example_kernel(a):
    tx = cuda.threadIdx.x
    bx = cuda.blockIdx.x
    bw = cuda.blockDim.x # 128 threads per block

    temp = cuda.shared.array(shape = (128 * 4), dtype=numba.float64)

    for i in range(4):
        temp[cuda.threadIdx.x + cuda.blockDim.x * i] = a[i]
\end{minted}
\caption{Optimised Code - Python}
\label{python-optimised-code}
\end{longlisting}

\subsection{Environment}
\label{environment}
The following are the run-time environment and hardware specifications used to execute the serial and GPU versions of the $\boldsymbol{q}$-LSKUM meshfree codes. 

\UseRawInputEncoding
\begin{lstlisting}[breaklines=true,numbers=none,basicstyle=\scriptsize]
SLURM_NODELIST=hsw222
SLURM_CHECKPOINT_IMAGE_DIR=/var/slurm/checkpoint
SLURM_JOB_NAME=bash
XDG_SESSION_ID=16625
SPACK_ROOT=/home/USER/spack
HOSTNAME=psgcluster
SLURM_TOPOLOGY_ADDR=hsw222
SLURMD_NODENAME=hsw222
SLURM_PRIO_PROCESS=0
SLURM_SRUN_COMM_PORT=35226
TERM=xterm-256color
SHELL=/bin/bash
HISTSIZE=1000
SLURM_PTY_WIN_ROW=64
SLURM_JOB_QOS=normal
XALT_EXECUTABLE_TRACKING=yes
SSH_CLIENT=10.40.205.31 44202 22
SLURM_TOPOLOGY_ADDR_PATTERN=node
TMPDIR=/tmp
LD_PRELOAD=/cm/shared/apps/xalt-gpu/xalt/xalt/lib64/libxalt_init.so
SINGULARITYENV_LD_PRELOAD=/cm/shared/apps/xalt-gpu/xalt/xalt/lib64/libxalt_init.so
QTDIR=/usr/lib64/qt-3.3
QTINC=/usr/lib64/qt-3.3/include
SSH_TTY=/dev/pts/112
QT_GRAPHICSSYSTEM_CHECKED=1
USER=USER
SLURM_NNODES=1
LS_COLORS=rs=0:di=38;5;27:ln=38;5;51:mh=44;38;5;15:pi=40;38;5;11:so=38;5;13:do=38;5;5:bd=48;5;232;38;5;11:cd=48;5;232;38;5;3:or=48;5;232;38;5;9:mi=05;48;5;232;38;5;15:su=48;5;196;38;5;15:sg=48;5;11;38;5;16:ca=48;5;196;38;5;226:tw=48;5;10;38;5;16:ow=48;5;10;38;5;21:st=48;5;21;38;5;15:ex=38;5;34:*.tar=38;5;9:*.tgz=38;5;9:*.arc=38;5;9:*.arj=38;5;9:*.taz=38;5;9:*.lha=38;5;9:*.lz4=38;5;9:*.lzh=38;5;9:*.lzma=38;5;9:*.tlz=38;5;9:*.txz=38;5;9:*.tzo=38;5;9:*.t7z=38;5;9:*.zip=38;5;9:*.z=38;5;9:*.Z=38;5;9:*.dz=38;5;9:*.gz=38;5;9:*.lrz=38;5;9:*.lz=38;5;9:*.lzo=38;5;9:*.xz=38;5;9:*.bz2=38;5;9:*.bz=38;5;9:*.tbz=38;5;9:*.tbz2=38;5;9:*.tz=38;5;9:*.deb=38;5;9:*.rpm=38;5;9:*.jar=38;5;9:*.war=38;5;9:*.ear=38;5;9:*.sar=38;5;9:*.rar=38;5;9:*.alz=38;5;9:*.ace=38;5;9:*.zoo=38;5;9:*.cpio=38;5;9:*.7z=38;5;9:*.rz=38;5;9:*.cab=38;5;9:*.jpg=38;5;13:*.jpeg=38;5;13:*.gif=38;5;13:*.bmp=38;5;13:*.pbm=38;5;13:*.pgm=38;5;13:*.ppm=38;5;13:*.tga=38;5;13:*.xbm=38;5;13:*.xpm=38;5;13:*.tif=38;5;13:*.tiff=38;5;13:*.png=38;5;13:*.svg=38;5;13:*.svgz=38;5;13:*.mng=38;5;13:*.pcx=38;5;13:*.mov=38;5;13:*.mpg=38;5;13:*.mpeg=38;5;13:*.m2v=38;5;13:*.mkv=38;5;13:*.webm=38;5;13:*.ogm=38;5;13:*.mp4=38;5;13:*.m4v=38;5;13:*.mp4v=38;5;13:*.vob=38;5;13:*.qt=38;5;13:*.nuv=38;5;13:*.wmv=38;5;13:*.asf=38;5;13:*.rm=38;5;13:*.rmvb=38;5;13:*.flc=38;5;13:*.avi=38;5;13:*.fli=38;5;13:*.flv=38;5;13:*.gl=38;5;13:*.dl=38;5;13:*.xcf=38;5;13:*.xwd=38;5;13:*.yuv=38;5;13:*.cgm=38;5;13:*.emf=38;5;13:*.axv=38;5;13:*.anx=38;5;13:*.ogv=38;5;13:*.ogx=38;5;13:*.aac=38;5;45:*.au=38;5;45:*.flac=38;5;45:*.mid=38;5;45:*.midi=38;5;45:*.mka=38;5;45:*.mp3=38;5;45:*.mpc=38;5;45:*.ogg=38;5;45:*.ra=38;5;45:*.wav=38;5;45:*.axa=38;5;45:*.oga=38;5;45:*.spx=38;5;45:*.xspf=38;5;45:
SLURM_STEP_NUM_NODES=1
SRUN_DEBUG=3
SLURM_JOBID=127191
SINGULARITY_CONTAINLIBS=/usr/lib64/libdcgm.so.1
SLURM_NTASKS=1
SLURM_LAUNCH_NODE_IPADDR=10.31.229.126
SLURM_STEP_ID=0
MAIL=/var/spool/mail/USER
PATH=/home/USER/spack/bin:/usr/lib64/qt-3.3/bin:/cm/shared/apps/xalt-gpu/xalt/xalt/bin:/usr/local/bin:/usr/bin:/usr/local/sbin:/usr/sbin:/opt/ibutils/bin:/sbin:/usr/sbin:/cm/extra/apps/Modules/3.2.10/bin:/home/USER/bin
SLURM_TASKS_PER_NODE=1
SLURM_STEP_LAUNCHER_PORT=35226
SLURM_WORKING_CLUSTER=psgcluster:psgcluster:6817:8192
_=/usr/bin/env
SLURM_JOB_ID=127191
PWD=/home/USER
SLURM_STEPID=0
SLURM_JOB_USER=USER
_LMFILES_=/cm/extra/ModuleFiles/Latest/xalt/2.6.3-gpu:/cm/extra/apps/Modules/3.2.10/Modules/modulefiles/null
SLURM_SRUN_COMM_HOST=10.31.229.126
LANG=en_US.UTF-8
MODULEPATH=/home/USER/spack/share/spack/modules/linux-centos7-nehalem:/home/USER/spack/share/spack/modules/linux-centos7-ivybridge::/cm/extra/apps/Modules/3.2.10/Modules/modulefiles:/cm/extra/eb/modules/all:/cm/extra/ModuleFiles/Latest:/cm/extra/ModuleFiles/Legacy:/cm/extra/Bundles:/cm/extra/PrgEnv/EZBuild/foss-2016a:/cm/extra/PrgEnv/EZBuild/foss-2016b:/cm/extra/PrgEnv/GCC+MPICH/2016-07-22:/cm/extra/PrgEnv/GCC+MVAPICH2/2016-10-26:/cm/extra/PrgEnv/GCC+OpenMPI/2016-04-16:/cm/extra/PrgEnv/GCC+OpenMPI/2016-06-24:/cm/extra/PrgEnv/GCC+OpenMPI/2016-10-12:/cm/extra/PrgEnv/GCC+OpenMPI/2016-12-08:/cm/extra/PrgEnv/GCC+OpenMPI/2017-10-05:/cm/extra/PrgEnv/GCC+OpenMPI/2017-11-06:/cm/extra/PrgEnv/GCC+OpenMPI/2018-05-24:/cm/extra/PrgEnv/Intel+IntelMPI/2017-03-02:/cm/extra/PrgEnv/Intel+MVAPICH2/2016-04-19:/cm/extra/PrgEnv/Intel+MVAPICH2/2016-10-26:/cm/extra/PrgEnv/Intel+OpenMPI/2016-03-21:/cm/extra/PrgEnv/Intel+OpenMPI/2016-08-03:/cm/extra/PrgEnv/Intel+OpenMPI/2017-02-16:/cm/extra/PrgEnv/PGI+MVAPICH2/2016-05-27:/cm/extra/PrgEnv/PGI+MVAPICH2/2017-11-05:/cm/extra/PrgEnv/PGI+MVAPICH2/2018-02-03:/cm/extra/PrgEnv/PGI+OpenMPI/2017-02-10:/cm/extra/PrgEnv/PGI+OpenMPI/2017-03-18:/cm/extra/PrgEnv/PGI+OpenMPI/2017-05-24:/cm/extra/PrgEnv/PGI+OpenMPI/2017-06-07:/cm/extra/PrgEnv/PGI+OpenMPI/2017-09-27:/cm/extra/PrgEnv/PGI+OpenMPI/2017-11-03:/cm/extra/PrgEnv/PGI+OpenMPI/2018-02-15:/cm/extra/PrgEnv/PGI+OpenMPI/2018-03-19:/cm/extra/PrgEnv/PGI+OpenMPI/2018-03-29:/cm/extra/PrgEnv/PGI+OpenMPI/2018-07-14:/cm/extra/PrgEnv/PGI+OpenMPI/2018-07-23:/cm/extra/PrgEnv/XL+OpenMPI/2017-07-26
SLURM_UMASK=0027
SLURM_PTY_WIN_COL=134
LOADEDMODULES=xalt/2.6.3-gpu:null
SLURM_JOB_UID=34783
SLURM_NODEID=0
SLURM_SUBMIT_DIR=/home/USER
XALT_SIGNAL_HANDLER=no
SLURM_NPROCS=1
SLURM_TASK_PID=47269
SLURM_CPUS_ON_NODE=64
SSH_ASKPASS=/usr/libexec/openssh/gnome-ssh-askpass
HISTCONTROL=ignoredups
SLURM_PROCID=0
SLURM_JOB_NODELIST=hsw222
SHLVL=3
HOME=/home/USER
SLURM_PTY_PORT=39797
SLURM_LOCALID=0
SLURM_CLUSTER_NAME=psgcluster
SLURM_JOB_CPUS_PER_NODE=64
SLURM_JOB_GID=34783
XALT_SCALAR_AND_SPSR_SAMPLING=yes
SLURM_SUBMIT_HOST=psgcluster
SLURM_GTIDS=0
SLURM_JOB_PARTITION=hsw_v100_32g
LOGNAME=USER
QTLIB=/usr/lib64/qt-3.3/lib
SINGULARITY_BINDPATH=/cm/shared/apps/xalt-gpu/xalt/xalt
SLURM_STEP_NUM_TASKS=1
SSH_CONNECTION=10.40.205.31 44202 10.31.115.254 22
SLURM_JOB_ACCOUNT=external
MODULESHOME=/cm/extra/apps/Modules/3.2.10
SLURM_JOB_NUM_NODES=1
COMPILER_PATH=/cm/shared/apps/xalt-gpu/xalt/xalt/bin
LESSOPEN=||/usr/bin/lesspipe.sh %s
SLURM_STEP_TASKS_PER_NODE=1
SPACK_PYTHON=/usr/bin/python3
SLURM_STEP_NODELIST=hsw222
XDG_RUNTIME_DIR=/run/user/34783
BASH_FUNC__spack_shell_wrapper()=() {  for var in LD_LIBRARY_PATH DYLD_LIBRARY_PATH DYLD_FALLBACK_LIBRARY_PATH;
 do
 eval "if [ -n \"\${${var}-}\" ]; then export SPACK_$var=\${${var}}; fi";
 done;
 if [ -n "${ZSH_VERSION:-}" ]; then
 emulate -L sh;
 fi;
 _sp_flags="";
 while [ ! -z ${1+x} ] && [ "${1#-}" != "${1}" ]; do
 _sp_flags="$_sp_flags $1";
 shift;
 done;
 if [ -n "$_sp_flags" ] && [ "${_sp_flags#*h}" != "${_sp_flags}" ] || [ "${_sp_flags#*V}" != "${_sp_flags}" ]; then
 command spack $_sp_flags "$@";
 return;
 fi;
 _sp_subcommand="";
 if [ ! -z ${1+x} ]; then
 _sp_subcommand="$1";
 shift;
 fi;
 case $_sp_subcommand in 
 "cd")
 _sp_arg="";
 if [ -n "$1" ]; then
 _sp_arg="$1";
 shift;
 fi;
 if [ "$_sp_arg" = "-h" ] || [ "$_sp_arg" = "--help" ]; then
 command spack cd -h;
 else
 LOC="$(spack location $_sp_arg "$@")";
 if [ -d "$LOC" ]; then
 cd "$LOC";
 else
 return 1;
 fi;
 fi;
 return
 ;;
 "env")
 _sp_arg="";
 if [ -n "$1" ]; then
 _sp_arg="$1";
 shift;
 fi;
 if [ "$_sp_arg" = "-h" ] || [ "$_sp_arg" = "--help" ]; then
 command spack env -h;
 else
 case $_sp_arg in 
 activate)
 _a=" $@";
 if [ -z ${1+x} ] || [ "${_a#* --sh}" != "$_a" ] || [ "${_a#* --csh}" != "$_a" ] || [ "${_a#* -h}" != "$_a" ] || [ "${_a#* --help}" != "$_a" ]; then
 command spack env activate "$@";
 else
 eval $(command spack $_sp_flags env activate --sh "$@");
 fi
 ;;
 deactivate)
 _a=" $@";
 if [ "${_a#* --sh}" != "$_a" ] || [ "${_a#* --csh}" != "$_a" ]; then
 command spack env deactivate "$@";
 else
 if [ -n "$*" ]; then
 command spack env deactivate -h;
 else
 eval $(command spack $_sp_flags env deactivate --sh);
 fi;
 fi
 ;;
 *)
 command spack env $_sp_arg "$@"
 ;;
 esac;
 fi;
 return
 ;;
 "load" | "unload")
 _a=" $@";
 if [ "${_a#* --sh}" != "$_a" ] || [ "${_a#* --csh}" != "$_a" ] || [ "${_a#* -h}" != "$_a" ] || [ "${_a#* --help}" != "$_a" ]; then
 command spack $_sp_flags $_sp_subcommand "$@";
 else
 eval $(command spack $_sp_flags $_sp_subcommand --sh "$@" ||                     echo "return 1");
 fi
 ;;
 *)
 command spack $_sp_flags $_sp_subcommand "$@"
 ;;
 esac
}
BASH_FUNC_module()=() {  eval `/cm/extra/apps/Modules/3.2.10/bin/modulecmd bash $*`
}
BASH_FUNC_spack()=() {  : this is a shell function from: /home/USER/spack/share/spack/setup-env.sh;
 : the real spack script is here: /home/USER/spack/bin/spack;
 _spack_shell_wrapper "$@";
 return $?
}
Linux hsw222 3.10.0-1062.4.1.el7.x86_64 #1 SMP Fri Oct 18 17:15:30 UTC 2019 x86_64 x86_64 x86_64 GNU/Linux
Architecture:          x86_64
CPU op-mode(s):        32-bit, 64-bit
Byte Order:            Little Endian
CPU(s):                64
On-line CPU(s) list:   0-31
Off-line CPU(s) list:  32-63
Thread(s) per core:    1
Core(s) per socket:    16
Socket(s):             2
NUMA node(s):          2
Vendor ID:             GenuineIntel
CPU family:            6
Model:                 63
Model name:            Intel(R) Xeon(R) CPU E5-2698 v3 @ 2.30GHz
Stepping:              2
CPU MHz:               2301.000
CPU max MHz:           2301.0000
CPU min MHz:           1200.0000
BogoMIPS:              4599.98
Virtualization:        VT-x
L1d cache:             32K
L1i cache:             32K
L2 cache:              256K
L3 cache:              40960K
NUMA node0 CPU(s):     0-15
NUMA node1 CPU(s):     16-31
Flags:                 fpu vme de pse tsc msr pae mce cx8 apic sep mtrr pge mca cmov pat pse36 clflush dts acpi mmx fxsr sse sse2 ss ht tm pbe syscall nx pdpe1gb rdtscp lm constant_tsc arch_perfmon pebs bts rep_good nopl xtopology nonstop_tsc aperfmperf eagerfpu pni pclmulqdq dtes64 monitor ds_cpl vmx smx est tm2 ssse3 sdbg fma cx16 xtpr pdcm pcid dca sse4_1 sse4_2 x2apic movbe popcnt xsave avx f16c rdrand lahf_lm abm epb invpcid_single intel_ppin ssbd ibrs ibpb tpr_shadow vnmi flexpriority ept vpid fsgsbase tsc_adjust bmi1 avx2 smep bmi2 erms invpcid cqm xsaveopt cqm_llc cqm_occup_llc dtherm ida arat pln pts md_clear
MemTotal:       264041332 kB
MemFree:        252255872 kB
MemAvailable:   250276344 kB
Buffers:             152 kB
Cached:          8568056 kB
SwapCached:            0 kB
Active:           389744 kB
Inactive:        8494016 kB
Active(anon):     379152 kB
Inactive(anon):  8454100 kB
Active(file):      10592 kB
Inactive(file):    39916 kB
Unevictable:           0 kB
Mlocked:               0 kB
SwapTotal:             0 kB
SwapFree:              0 kB
Dirty:                 0 kB
Writeback:             0 kB
AnonPages:        315680 kB
Mapped:            96096 kB
Shmem:           8517700 kB
Slab:             565848 kB
SReclaimable:     183324 kB
SUnreclaim:       382524 kB
KernelStack:       12688 kB
PageTables:         7992 kB
NFS_Unstable:          0 kB
Bounce:                0 kB
WritebackTmp:          0 kB
CommitLimit:    132020664 kB
Committed_AS:   10643848 kB
VmallocTotal:   34359738367 kB
VmallocUsed:      866172 kB
VmallocChunk:   34224519164 kB
HardwareCorrupted:     0 kB
AnonHugePages:    129024 kB
CmaTotal:              0 kB
CmaFree:               0 kB
HugePages_Total:       0
HugePages_Free:        0
HugePages_Rsvd:        0
HugePages_Surp:        0
Hugepagesize:       2048 kB
DirectMap4k:     8410832 kB
DirectMap2M:    250480640 kB
DirectMap1G:    11534336 kB
NAME   MAJ:MIN RM   SIZE RO TYPE MOUNTPOINT
sda      8:0    0 558.9G  0 disk 
├─sda1   8:1    0   7.7G  0 part 
└─sda2   8:2    0 551.3G  0 part /local
loop0    7:0    0         1 loop 
loop1    7:1    0         1 loop 
loop2    7:2    0         1 loop 
loop3    7:3    0         1 loop 
loop4    7:4    0         1 loop 
loop5    7:5    0         1 loop 
loop6    7:6    0         1 loop 
loop7    7:7    0         1 loop 
loop8    7:8    0         1 loop 
loop9    7:9    0         1 loop 
loop10   7:10   0         1 loop 
loop11   7:11   0         1 loop 
loop12   7:12   0         1 loop 
loop13   7:13   0         1 loop 
loop14   7:14   0         1 loop 
loop15   7:15   0         1 loop 
loop16   7:16   0         1 loop 
loop17   7:17   0         1 loop 
loop18   7:18   0         1 loop 
loop19   7:19   0         1 loop 
[0:0:0:0]    disk    ATA      INTEL SSDSC2BB60 0370  /dev/sda    600GB
Sun Jul 18 11:01:41 2021       
+-----------------------------------------------------------------------------+
| NVIDIA-SMI 460.73.01    Driver Version: 460.73.01    CUDA Version: 11.2     |
|-------------------------------+----------------------+----------------------+
| GPU  Name        Persistence-M| Bus-Id        Disp.A | Volatile Uncorr. ECC |
| Fan  Temp  Perf  Pwr:Usage/Cap|         Memory-Usage | GPU-Util  Compute M. |
|                               |                      |               MIG M. |
|===============================+======================+======================|
|   0  Tesla V100-PCIE...  On   | 00000000:05:00.0 Off |                    0 |
| N/A   29C    P0    24W / 250W |      0MiB / 32510MiB |      0%      Default |
|                               |                      |                  N/A |
+-------------------------------+----------------------+----------------------+
|   1  Tesla V100-PCIE...  On   | 00000000:06:00.0 Off |                    0 |
| N/A   31C    P0    25W / 250W |      0MiB / 32510MiB |      0%      Default |
|                               |                      |                  N/A |
+-------------------------------+----------------------+----------------------+
|   2  Tesla V100-PCIE...  On   | 00000000:84:00.0 Off |                    0 |
| N/A   30C    P0    24W / 250W |      0MiB / 32510MiB |      0%      Default |
|                               |                      |                  N/A |
+-------------------------------+----------------------+----------------------+
|   3  Tesla V100-PCIE...  On   | 00000000:85:00.0 Off |                    0 |
| N/A   32C    P0    24W / 250W |      0MiB / 32510MiB |      0%      Default |
|                               |                      |                  N/A |
+-------------------------------+----------------------+----------------------+
                                                                               
+-----------------------------------------------------------------------------+
| Processes:                                                                  |
|  GPU   GI   CI        PID   Type   Process name                  GPU Memory |
|        ID   ID                                                   Usage      |
|=============================================================================|
|  No running processes found                                                 |
+-----------------------------------------------------------------------------+
00:00.0 Host bridge: Intel Corporation Xeon E7 v3/Xeon E5 v3/Core i7 DMI2 (rev 02)
00:01.0 PCI bridge: Intel Corporation Xeon E7 v3/Xeon E5 v3/Core i7 PCI Express Root Port 1 (rev 02)
00:02.0 PCI bridge: Intel Corporation Xeon E7 v3/Xeon E5 v3/Core i7 PCI Express Root Port 2 (rev 02)
00:03.0 PCI bridge: Intel Corporation Xeon E7 v3/Xeon E5 v3/Core i7 PCI Express Root Port 3 (rev 02)
00:04.0 System peripheral: Intel Corporation Xeon E7 v3/Xeon E5 v3/Core i7 DMA Channel 0 (rev 02)
00:04.1 System peripheral: Intel Corporation Xeon E7 v3/Xeon E5 v3/Core i7 DMA Channel 1 (rev 02)
00:04.2 System peripheral: Intel Corporation Xeon E7 v3/Xeon E5 v3/Core i7 DMA Channel 2 (rev 02)
00:04.3 System peripheral: Intel Corporation Xeon E7 v3/Xeon E5 v3/Core i7 DMA Channel 3 (rev 02)
00:04.4 System peripheral: Intel Corporation Xeon E7 v3/Xeon E5 v3/Core i7 DMA Channel 4 (rev 02)
00:04.5 System peripheral: Intel Corporation Xeon E7 v3/Xeon E5 v3/Core i7 DMA Channel 5 (rev 02)
00:04.6 System peripheral: Intel Corporation Xeon E7 v3/Xeon E5 v3/Core i7 DMA Channel 6 (rev 02)
00:04.7 System peripheral: Intel Corporation Xeon E7 v3/Xeon E5 v3/Core i7 DMA Channel 7 (rev 02)
00:05.0 System peripheral: Intel Corporation Xeon E7 v3/Xeon E5 v3/Core i7 Address Map, VTd_Misc, System Management (rev 02)
00:05.1 System peripheral: Intel Corporation Xeon E7 v3/Xeon E5 v3/Core i7 Hot Plug (rev 02)
00:05.2 System peripheral: Intel Corporation Xeon E7 v3/Xeon E5 v3/Core i7 RAS, Control Status and Global Errors (rev 02)
00:05.4 PIC: Intel Corporation Xeon E7 v3/Xeon E5 v3/Core i7 I/O APIC (rev 02)
00:11.0 Unassigned class [ff00]: Intel Corporation C610/X99 series chipset SPSR (rev 05)
00:11.4 SATA controller: Intel Corporation C610/X99 series chipset sSATA Controller [AHCI mode] (rev 05)
00:14.0 USB controller: Intel Corporation C610/X99 series chipset USB xHCI Host Controller (rev 05)
00:16.0 Communication controller: Intel Corporation C610/X99 series chipset MEI Controller #1 (rev 05)
00:16.1 Communication controller: Intel Corporation C610/X99 series chipset MEI Controller #2 (rev 05)
00:1a.0 USB controller: Intel Corporation C610/X99 series chipset USB Enhanced Host Controller #2 (rev 05)
00:1c.0 PCI bridge: Intel Corporation C610/X99 series chipset PCI Express Root Port #1 (rev d5)
00:1c.4 PCI bridge: Intel Corporation C610/X99 series chipset PCI Express Root Port #5 (rev d5)
00:1d.0 USB controller: Intel Corporation C610/X99 series chipset USB Enhanced Host Controller #1 (rev 05)
00:1f.0 ISA bridge: Intel Corporation C610/X99 series chipset LPC Controller (rev 05)
00:1f.2 SATA controller: Intel Corporation C610/X99 series chipset 6-Port SATA Controller [AHCI mode] (rev 05)
00:1f.3 SMBus: Intel Corporation C610/X99 series chipset SMBus Controller (rev 05)
01:00.0 Ethernet controller: Intel Corporation Ethernet Controller 10-Gigabit X540-AT2 (rev 01)
01:00.1 Ethernet controller: Intel Corporation Ethernet Controller 10-Gigabit X540-AT2 (rev 01)
03:00.0 PCI bridge: PLX Technology, Inc. PEX 8747 48-Lane, 5-Port PCI Express Gen 3 (8.0 GT/s) Switch (rev ca)
04:08.0 PCI bridge: PLX Technology, Inc. PEX 8747 48-Lane, 5-Port PCI Express Gen 3 (8.0 GT/s) Switch (rev ca)
04:10.0 PCI bridge: PLX Technology, Inc. PEX 8747 48-Lane, 5-Port PCI Express Gen 3 (8.0 GT/s) Switch (rev ca)
05:00.0 3D controller: NVIDIA Corporation GV100GL [Tesla V100 PCIe 32GB] (rev a1)
06:00.0 3D controller: NVIDIA Corporation GV100GL [Tesla V100 PCIe 32GB] (rev a1)
08:00.0 PCI bridge: ASPEED Technology, Inc. AST1150 PCI-to-PCI Bridge (rev 03)
09:00.0 VGA compatible controller: ASPEED Technology, Inc. ASPEED Graphics Family (rev 30)
7f:08.0 System peripheral: Intel Corporation Xeon E7 v3/Xeon E5 v3/Core i7 QPI Link 0 (rev 02)
7f:08.2 Performance counters: Intel Corporation Xeon E7 v3/Xeon E5 v3/Core i7 QPI Link 0 (rev 02)
7f:08.3 System peripheral: Intel Corporation Xeon E7 v3/Xeon E5 v3/Core i7 QPI Link 0 (rev 02)
7f:09.0 System peripheral: Intel Corporation Xeon E7 v3/Xeon E5 v3/Core i7 QPI Link 1 (rev 02)
7f:09.2 Performance counters: Intel Corporation Xeon E7 v3/Xeon E5 v3/Core i7 QPI Link 1 (rev 02)
7f:09.3 System peripheral: Intel Corporation Xeon E7 v3/Xeon E5 v3/Core i7 QPI Link 1 (rev 02)
7f:0b.0 System peripheral: Intel Corporation Xeon E7 v3/Xeon E5 v3/Core i7 R3 QPI Link 0 & 1 Monitoring (rev 02)
7f:0b.1 Performance counters: Intel Corporation Xeon E7 v3/Xeon E5 v3/Core i7 R3 QPI Link 0 & 1 Monitoring (rev 02)
7f:0b.2 Performance counters: Intel Corporation Xeon E7 v3/Xeon E5 v3/Core i7 R3 QPI Link 0 & 1 Monitoring (rev 02)
7f:0c.0 System peripheral: Intel Corporation Xeon E7 v3/Xeon E5 v3/Core i7 Unicast Registers (rev 02)
7f:0c.1 System peripheral: Intel Corporation Xeon E7 v3/Xeon E5 v3/Core i7 Unicast Registers (rev 02)
7f:0c.2 System peripheral: Intel Corporation Xeon E7 v3/Xeon E5 v3/Core i7 Unicast Registers (rev 02)
7f:0c.3 System peripheral: Intel Corporation Xeon E7 v3/Xeon E5 v3/Core i7 Unicast Registers (rev 02)
7f:0c.4 System peripheral: Intel Corporation Xeon E7 v3/Xeon E5 v3/Core i7 Unicast Registers (rev 02)
7f:0c.5 System peripheral: Intel Corporation Xeon E7 v3/Xeon E5 v3/Core i7 Unicast Registers (rev 02)
7f:0c.6 System peripheral: Intel Corporation Xeon E7 v3/Xeon E5 v3/Core i7 Unicast Registers (rev 02)
7f:0c.7 System peripheral: Intel Corporation Xeon E7 v3/Xeon E5 v3/Core i7 Unicast Registers (rev 02)
7f:0d.0 System peripheral: Intel Corporation Xeon E7 v3/Xeon E5 v3/Core i7 Unicast Registers (rev 02)
7f:0d.1 System peripheral: Intel Corporation Xeon E7 v3/Xeon E5 v3/Core i7 Unicast Registers (rev 02)
7f:0d.2 System peripheral: Intel Corporation Xeon E7 v3/Xeon E5 v3/Core i7 Unicast Registers (rev 02)
7f:0d.3 System peripheral: Intel Corporation Xeon E7 v3/Xeon E5 v3/Core i7 Unicast Registers (rev 02)
7f:0d.4 System peripheral: Intel Corporation Xeon E7 v3/Xeon E5 v3/Core i7 Unicast Registers (rev 02)
7f:0d.5 System peripheral: Intel Corporation Xeon E7 v3/Xeon E5 v3/Core i7 Unicast Registers (rev 02)
7f:0d.6 System peripheral: Intel Corporation Xeon E7 v3/Xeon E5 v3/Core i7 Unicast Registers (rev 02)
7f:0d.7 System peripheral: Intel Corporation Xeon E7 v3/Xeon E5 v3/Core i7 Unicast Registers (rev 02)
7f:0f.0 System peripheral: Intel Corporation Xeon E7 v3/Xeon E5 v3/Core i7 Buffered Ring Agent (rev 02)
7f:0f.1 System peripheral: Intel Corporation Xeon E7 v3/Xeon E5 v3/Core i7 Buffered Ring Agent (rev 02)
7f:0f.2 System peripheral: Intel Corporation Xeon E7 v3/Xeon E5 v3/Core i7 Buffered Ring Agent (rev 02)
7f:0f.3 System peripheral: Intel Corporation Xeon E7 v3/Xeon E5 v3/Core i7 Buffered Ring Agent (rev 02)
7f:0f.4 System peripheral: Intel Corporation Xeon E7 v3/Xeon E5 v3/Core i7 System Address Decoder & Broadcast Registers (rev 02)
7f:0f.5 System peripheral: Intel Corporation Xeon E7 v3/Xeon E5 v3/Core i7 System Address Decoder & Broadcast Registers (rev 02)
7f:0f.6 System peripheral: Intel Corporation Xeon E7 v3/Xeon E5 v3/Core i7 System Address Decoder & Broadcast Registers (rev 02)
7f:10.0 System peripheral: Intel Corporation Xeon E7 v3/Xeon E5 v3/Core i7 PCIe Ring Interface (rev 02)
7f:10.1 Performance counters: Intel Corporation Xeon E7 v3/Xeon E5 v3/Core i7 PCIe Ring Interface (rev 02)
7f:10.5 System peripheral: Intel Corporation Xeon E7 v3/Xeon E5 v3/Core i7 Scratchpad & Semaphore Registers (rev 02)
7f:10.6 Performance counters: Intel Corporation Xeon E7 v3/Xeon E5 v3/Core i7 Scratchpad & Semaphore Registers (rev 02)
7f:10.7 System peripheral: Intel Corporation Xeon E7 v3/Xeon E5 v3/Core i7 Scratchpad & Semaphore Registers (rev 02)
7f:12.0 System peripheral: Intel Corporation Xeon E7 v3/Xeon E5 v3/Core i7 Home Agent 0 (rev 02)
7f:12.1 Performance counters: Intel Corporation Xeon E7 v3/Xeon E5 v3/Core i7 Home Agent 0 (rev 02)
7f:12.4 System peripheral: Intel Corporation Xeon E7 v3/Xeon E5 v3/Core i7 Home Agent 1 (rev 02)
7f:12.5 Performance counters: Intel Corporation Xeon E7 v3/Xeon E5 v3/Core i7 Home Agent 1 (rev 02)
7f:13.0 System peripheral: Intel Corporation Xeon E7 v3/Xeon E5 v3/Core i7 Integrated Memory Controller 0 Target Address, Thermal & RAS Registers (rev 02)
7f:13.1 System peripheral: Intel Corporation Xeon E7 v3/Xeon E5 v3/Core i7 Integrated Memory Controller 0 Target Address, Thermal & RAS Registers (rev 02)
7f:13.2 System peripheral: Intel Corporation Xeon E7 v3/Xeon E5 v3/Core i7 Integrated Memory Controller 0 Channel Target Address Decoder (rev 02)
7f:13.3 System peripheral: Intel Corporation Xeon E7 v3/Xeon E5 v3/Core i7 Integrated Memory Controller 0 Channel Target Address Decoder (rev 02)
7f:13.6 System peripheral: Intel Corporation Xeon E7 v3/Xeon E5 v3/Core i7 DDRIO Channel 0/1 Broadcast (rev 02)
7f:13.7 System peripheral: Intel Corporation Xeon E7 v3/Xeon E5 v3/Core i7 DDRIO Global Broadcast (rev 02)
7f:14.0 System peripheral: Intel Corporation Xeon E7 v3/Xeon E5 v3/Core i7 Integrated Memory Controller 0 Channel 0 Thermal Control (rev 02)
7f:14.1 System peripheral: Intel Corporation Xeon E7 v3/Xeon E5 v3/Core i7 Integrated Memory Controller 0 Channel 1 Thermal Control (rev 02)
7f:14.2 System peripheral: Intel Corporation Xeon E7 v3/Xeon E5 v3/Core i7 Integrated Memory Controller 0 Channel 0 ERROR Registers (rev 02)
7f:14.3 System peripheral: Intel Corporation Xeon E7 v3/Xeon E5 v3/Core i7 Integrated Memory Controller 0 Channel 1 ERROR Registers (rev 02)
7f:14.4 System peripheral: Intel Corporation Xeon E7 v3/Xeon E5 v3/Core i7 DDRIO (VMSE) 0 & 1 (rev 02)
7f:14.5 System peripheral: Intel Corporation Xeon E7 v3/Xeon E5 v3/Core i7 DDRIO (VMSE) 0 & 1 (rev 02)
7f:14.6 System peripheral: Intel Corporation Xeon E7 v3/Xeon E5 v3/Core i7 DDRIO (VMSE) 0 & 1 (rev 02)
7f:14.7 System peripheral: Intel Corporation Xeon E7 v3/Xeon E5 v3/Core i7 DDRIO (VMSE) 0 & 1 (rev 02)
7f:16.0 System peripheral: Intel Corporation Xeon E7 v3/Xeon E5 v3/Core i7 Integrated Memory Controller 1 Target Address, Thermal & RAS Registers (rev 02)
7f:16.1 System peripheral: Intel Corporation Xeon E7 v3/Xeon E5 v3/Core i7 Integrated Memory Controller 1 Target Address, Thermal & RAS Registers (rev 02)
7f:16.2 System peripheral: Intel Corporation Xeon E7 v3/Xeon E5 v3/Core i7 Integrated Memory Controller 1 Channel Target Address Decoder (rev 02)
7f:16.3 System peripheral: Intel Corporation Xeon E7 v3/Xeon E5 v3/Core i7 Integrated Memory Controller 1 Channel Target Address Decoder (rev 02)
7f:16.6 System peripheral: Intel Corporation Xeon E7 v3/Xeon E5 v3/Core i7 DDRIO Channel 2/3 Broadcast (rev 02)
7f:16.7 System peripheral: Intel Corporation Xeon E7 v3/Xeon E5 v3/Core i7 DDRIO Global Broadcast (rev 02)
7f:17.0 System peripheral: Intel Corporation Xeon E7 v3/Xeon E5 v3/Core i7 Integrated Memory Controller 1 Channel 0 Thermal Control (rev 02)
7f:17.1 System peripheral: Intel Corporation Xeon E7 v3/Xeon E5 v3/Core i7 Integrated Memory Controller 1 Channel 1 Thermal Control (rev 02)
7f:17.2 System peripheral: Intel Corporation Xeon E7 v3/Xeon E5 v3/Core i7 Integrated Memory Controller 1 Channel 0 ERROR Registers (rev 02)
7f:17.3 System peripheral: Intel Corporation Xeon E7 v3/Xeon E5 v3/Core i7 Integrated Memory Controller 1 Channel 1 ERROR Registers (rev 02)
7f:17.4 System peripheral: Intel Corporation Xeon E7 v3/Xeon E5 v3/Core i7 DDRIO (VMSE) 2 & 3 (rev 02)
7f:17.5 System peripheral: Intel Corporation Xeon E7 v3/Xeon E5 v3/Core i7 DDRIO (VMSE) 2 & 3 (rev 02)
7f:17.6 System peripheral: Intel Corporation Xeon E7 v3/Xeon E5 v3/Core i7 DDRIO (VMSE) 2 & 3 (rev 02)
7f:17.7 System peripheral: Intel Corporation Xeon E7 v3/Xeon E5 v3/Core i7 DDRIO (VMSE) 2 & 3 (rev 02)
7f:1e.0 System peripheral: Intel Corporation Xeon E7 v3/Xeon E5 v3/Core i7 Power Control Unit (rev 02)
7f:1e.1 System peripheral: Intel Corporation Xeon E7 v3/Xeon E5 v3/Core i7 Power Control Unit (rev 02)
7f:1e.2 System peripheral: Intel Corporation Xeon E7 v3/Xeon E5 v3/Core i7 Power Control Unit (rev 02)
7f:1e.3 System peripheral: Intel Corporation Xeon E7 v3/Xeon E5 v3/Core i7 Power Control Unit (rev 02)
7f:1e.4 System peripheral: Intel Corporation Xeon E7 v3/Xeon E5 v3/Core i7 Power Control Unit (rev 02)
7f:1f.0 System peripheral: Intel Corporation Xeon E7 v3/Xeon E5 v3/Core i7 VCU (rev 02)
7f:1f.2 System peripheral: Intel Corporation Xeon E7 v3/Xeon E5 v3/Core i7 VCU (rev 02)
80:01.0 PCI bridge: Intel Corporation Xeon E7 v3/Xeon E5 v3/Core i7 PCI Express Root Port 1 (rev 02)
80:03.0 PCI bridge: Intel Corporation Xeon E7 v3/Xeon E5 v3/Core i7 PCI Express Root Port 3 (rev 02)
80:04.0 System peripheral: Intel Corporation Xeon E7 v3/Xeon E5 v3/Core i7 DMA Channel 0 (rev 02)
80:04.1 System peripheral: Intel Corporation Xeon E7 v3/Xeon E5 v3/Core i7 DMA Channel 1 (rev 02)
80:04.2 System peripheral: Intel Corporation Xeon E7 v3/Xeon E5 v3/Core i7 DMA Channel 2 (rev 02)
80:04.3 System peripheral: Intel Corporation Xeon E7 v3/Xeon E5 v3/Core i7 DMA Channel 3 (rev 02)
80:04.4 System peripheral: Intel Corporation Xeon E7 v3/Xeon E5 v3/Core i7 DMA Channel 4 (rev 02)
80:04.5 System peripheral: Intel Corporation Xeon E7 v3/Xeon E5 v3/Core i7 DMA Channel 5 (rev 02)
80:04.6 System peripheral: Intel Corporation Xeon E7 v3/Xeon E5 v3/Core i7 DMA Channel 6 (rev 02)
80:04.7 System peripheral: Intel Corporation Xeon E7 v3/Xeon E5 v3/Core i7 DMA Channel 7 (rev 02)
80:05.0 System peripheral: Intel Corporation Xeon E7 v3/Xeon E5 v3/Core i7 Address Map, VTd_Misc, System Management (rev 02)
80:05.1 System peripheral: Intel Corporation Xeon E7 v3/Xeon E5 v3/Core i7 Hot Plug (rev 02)
80:05.2 System peripheral: Intel Corporation Xeon E7 v3/Xeon E5 v3/Core i7 RAS, Control Status and Global Errors (rev 02)
80:05.4 PIC: Intel Corporation Xeon E7 v3/Xeon E5 v3/Core i7 I/O APIC (rev 02)
81:00.0 Infiniband controller: Mellanox Technologies MT27600 [Connect-IB]
82:00.0 PCI bridge: PLX Technology, Inc. PEX 8747 48-Lane, 5-Port PCI Express Gen 3 (8.0 GT/s) Switch (rev ca)
83:08.0 PCI bridge: PLX Technology, Inc. PEX 8747 48-Lane, 5-Port PCI Express Gen 3 (8.0 GT/s) Switch (rev ca)
83:10.0 PCI bridge: PLX Technology, Inc. PEX 8747 48-Lane, 5-Port PCI Express Gen 3 (8.0 GT/s) Switch (rev ca)
84:00.0 3D controller: NVIDIA Corporation GV100GL [Tesla V100 PCIe 32GB] (rev a1)
85:00.0 3D controller: NVIDIA Corporation GV100GL [Tesla V100 PCIe 32GB] (rev a1)
ff:08.0 System peripheral: Intel Corporation Xeon E7 v3/Xeon E5 v3/Core i7 QPI Link 0 (rev 02)
ff:08.2 Performance counters: Intel Corporation Xeon E7 v3/Xeon E5 v3/Core i7 QPI Link 0 (rev 02)
ff:08.3 System peripheral: Intel Corporation Xeon E7 v3/Xeon E5 v3/Core i7 QPI Link 0 (rev 02)
ff:09.0 System peripheral: Intel Corporation Xeon E7 v3/Xeon E5 v3/Core i7 QPI Link 1 (rev 02)
ff:09.2 Performance counters: Intel Corporation Xeon E7 v3/Xeon E5 v3/Core i7 QPI Link 1 (rev 02)
ff:09.3 System peripheral: Intel Corporation Xeon E7 v3/Xeon E5 v3/Core i7 QPI Link 1 (rev 02)
ff:0b.0 System peripheral: Intel Corporation Xeon E7 v3/Xeon E5 v3/Core i7 R3 QPI Link 0 & 1 Monitoring (rev 02)
ff:0b.1 Performance counters: Intel Corporation Xeon E7 v3/Xeon E5 v3/Core i7 R3 QPI Link 0 & 1 Monitoring (rev 02)
ff:0b.2 Performance counters: Intel Corporation Xeon E7 v3/Xeon E5 v3/Core i7 R3 QPI Link 0 & 1 Monitoring (rev 02)
ff:0c.0 System peripheral: Intel Corporation Xeon E7 v3/Xeon E5 v3/Core i7 Unicast Registers (rev 02)
ff:0c.1 System peripheral: Intel Corporation Xeon E7 v3/Xeon E5 v3/Core i7 Unicast Registers (rev 02)
ff:0c.2 System peripheral: Intel Corporation Xeon E7 v3/Xeon E5 v3/Core i7 Unicast Registers (rev 02)
ff:0c.3 System peripheral: Intel Corporation Xeon E7 v3/Xeon E5 v3/Core i7 Unicast Registers (rev 02)
ff:0c.4 System peripheral: Intel Corporation Xeon E7 v3/Xeon E5 v3/Core i7 Unicast Registers (rev 02)
ff:0c.5 System peripheral: Intel Corporation Xeon E7 v3/Xeon E5 v3/Core i7 Unicast Registers (rev 02)
ff:0c.6 System peripheral: Intel Corporation Xeon E7 v3/Xeon E5 v3/Core i7 Unicast Registers (rev 02)
ff:0c.7 System peripheral: Intel Corporation Xeon E7 v3/Xeon E5 v3/Core i7 Unicast Registers (rev 02)
ff:0d.0 System peripheral: Intel Corporation Xeon E7 v3/Xeon E5 v3/Core i7 Unicast Registers (rev 02)
ff:0d.1 System peripheral: Intel Corporation Xeon E7 v3/Xeon E5 v3/Core i7 Unicast Registers (rev 02)
ff:0d.2 System peripheral: Intel Corporation Xeon E7 v3/Xeon E5 v3/Core i7 Unicast Registers (rev 02)
ff:0d.3 System peripheral: Intel Corporation Xeon E7 v3/Xeon E5 v3/Core i7 Unicast Registers (rev 02)
ff:0d.4 System peripheral: Intel Corporation Xeon E7 v3/Xeon E5 v3/Core i7 Unicast Registers (rev 02)
ff:0d.5 System peripheral: Intel Corporation Xeon E7 v3/Xeon E5 v3/Core i7 Unicast Registers (rev 02)
ff:0d.6 System peripheral: Intel Corporation Xeon E7 v3/Xeon E5 v3/Core i7 Unicast Registers (rev 02)
ff:0d.7 System peripheral: Intel Corporation Xeon E7 v3/Xeon E5 v3/Core i7 Unicast Registers (rev 02)
ff:0f.0 System peripheral: Intel Corporation Xeon E7 v3/Xeon E5 v3/Core i7 Buffered Ring Agent (rev 02)
ff:0f.1 System peripheral: Intel Corporation Xeon E7 v3/Xeon E5 v3/Core i7 Buffered Ring Agent (rev 02)
ff:0f.2 System peripheral: Intel Corporation Xeon E7 v3/Xeon E5 v3/Core i7 Buffered Ring Agent (rev 02)
ff:0f.3 System peripheral: Intel Corporation Xeon E7 v3/Xeon E5 v3/Core i7 Buffered Ring Agent (rev 02)
ff:0f.4 System peripheral: Intel Corporation Xeon E7 v3/Xeon E5 v3/Core i7 System Address Decoder & Broadcast Registers (rev 02)
ff:0f.5 System peripheral: Intel Corporation Xeon E7 v3/Xeon E5 v3/Core i7 System Address Decoder & Broadcast Registers (rev 02)
ff:0f.6 System peripheral: Intel Corporation Xeon E7 v3/Xeon E5 v3/Core i7 System Address Decoder & Broadcast Registers (rev 02)
ff:10.0 System peripheral: Intel Corporation Xeon E7 v3/Xeon E5 v3/Core i7 PCIe Ring Interface (rev 02)
ff:10.1 Performance counters: Intel Corporation Xeon E7 v3/Xeon E5 v3/Core i7 PCIe Ring Interface (rev 02)
ff:10.5 System peripheral: Intel Corporation Xeon E7 v3/Xeon E5 v3/Core i7 Scratchpad & Semaphore Registers (rev 02)
ff:10.6 Performance counters: Intel Corporation Xeon E7 v3/Xeon E5 v3/Core i7 Scratchpad & Semaphore Registers (rev 02)
ff:10.7 System peripheral: Intel Corporation Xeon E7 v3/Xeon E5 v3/Core i7 Scratchpad & Semaphore Registers (rev 02)
ff:12.0 System peripheral: Intel Corporation Xeon E7 v3/Xeon E5 v3/Core i7 Home Agent 0 (rev 02)
ff:12.1 Performance counters: Intel Corporation Xeon E7 v3/Xeon E5 v3/Core i7 Home Agent 0 (rev 02)
ff:12.4 System peripheral: Intel Corporation Xeon E7 v3/Xeon E5 v3/Core i7 Home Agent 1 (rev 02)
ff:12.5 Performance counters: Intel Corporation Xeon E7 v3/Xeon E5 v3/Core i7 Home Agent 1 (rev 02)
ff:13.0 System peripheral: Intel Corporation Xeon E7 v3/Xeon E5 v3/Core i7 Integrated Memory Controller 0 Target Address, Thermal & RAS Registers (rev 02)
ff:13.1 System peripheral: Intel Corporation Xeon E7 v3/Xeon E5 v3/Core i7 Integrated Memory Controller 0 Target Address, Thermal & RAS Registers (rev 02)
ff:13.2 System peripheral: Intel Corporation Xeon E7 v3/Xeon E5 v3/Core i7 Integrated Memory Controller 0 Channel Target Address Decoder (rev 02)
ff:13.3 System peripheral: Intel Corporation Xeon E7 v3/Xeon E5 v3/Core i7 Integrated Memory Controller 0 Channel Target Address Decoder (rev 02)
ff:13.6 System peripheral: Intel Corporation Xeon E7 v3/Xeon E5 v3/Core i7 DDRIO Channel 0/1 Broadcast (rev 02)
ff:13.7 System peripheral: Intel Corporation Xeon E7 v3/Xeon E5 v3/Core i7 DDRIO Global Broadcast (rev 02)
ff:14.0 System peripheral: Intel Corporation Xeon E7 v3/Xeon E5 v3/Core i7 Integrated Memory Controller 0 Channel 0 Thermal Control (rev 02)
ff:14.1 System peripheral: Intel Corporation Xeon E7 v3/Xeon E5 v3/Core i7 Integrated Memory Controller 0 Channel 1 Thermal Control (rev 02)
ff:14.2 System peripheral: Intel Corporation Xeon E7 v3/Xeon E5 v3/Core i7 Integrated Memory Controller 0 Channel 0 ERROR Registers (rev 02)
ff:14.3 System peripheral: Intel Corporation Xeon E7 v3/Xeon E5 v3/Core i7 Integrated Memory Controller 0 Channel 1 ERROR Registers (rev 02)
ff:14.4 System peripheral: Intel Corporation Xeon E7 v3/Xeon E5 v3/Core i7 DDRIO (VMSE) 0 & 1 (rev 02)
ff:14.5 System peripheral: Intel Corporation Xeon E7 v3/Xeon E5 v3/Core i7 DDRIO (VMSE) 0 & 1 (rev 02)
ff:14.6 System peripheral: Intel Corporation Xeon E7 v3/Xeon E5 v3/Core i7 DDRIO (VMSE) 0 & 1 (rev 02)
ff:14.7 System peripheral: Intel Corporation Xeon E7 v3/Xeon E5 v3/Core i7 DDRIO (VMSE) 0 & 1 (rev 02)
ff:16.0 System peripheral: Intel Corporation Xeon E7 v3/Xeon E5 v3/Core i7 Integrated Memory Controller 1 Target Address, Thermal & RAS Registers (rev 02)
ff:16.1 System peripheral: Intel Corporation Xeon E7 v3/Xeon E5 v3/Core i7 Integrated Memory Controller 1 Target Address, Thermal & RAS Registers (rev 02)
ff:16.2 System peripheral: Intel Corporation Xeon E7 v3/Xeon E5 v3/Core i7 Integrated Memory Controller 1 Channel Target Address Decoder (rev 02)
ff:16.3 System peripheral: Intel Corporation Xeon E7 v3/Xeon E5 v3/Core i7 Integrated Memory Controller 1 Channel Target Address Decoder (rev 02)
ff:16.6 System peripheral: Intel Corporation Xeon E7 v3/Xeon E5 v3/Core i7 DDRIO Channel 2/3 Broadcast (rev 02)
ff:16.7 System peripheral: Intel Corporation Xeon E7 v3/Xeon E5 v3/Core i7 DDRIO Global Broadcast (rev 02)
ff:17.0 System peripheral: Intel Corporation Xeon E7 v3/Xeon E5 v3/Core i7 Integrated Memory Controller 1 Channel 0 Thermal Control (rev 02)
ff:17.1 System peripheral: Intel Corporation Xeon E7 v3/Xeon E5 v3/Core i7 Integrated Memory Controller 1 Channel 1 Thermal Control (rev 02)
ff:17.2 System peripheral: Intel Corporation Xeon E7 v3/Xeon E5 v3/Core i7 Integrated Memory Controller 1 Channel 0 ERROR Registers (rev 02)
ff:17.3 System peripheral: Intel Corporation Xeon E7 v3/Xeon E5 v3/Core i7 Integrated Memory Controller 1 Channel 1 ERROR Registers (rev 02)
ff:17.4 System peripheral: Intel Corporation Xeon E7 v3/Xeon E5 v3/Core i7 DDRIO (VMSE) 2 & 3 (rev 02)
ff:17.5 System peripheral: Intel Corporation Xeon E7 v3/Xeon E5 v3/Core i7 DDRIO (VMSE) 2 & 3 (rev 02)
ff:17.6 System peripheral: Intel Corporation Xeon E7 v3/Xeon E5 v3/Core i7 DDRIO (VMSE) 2 & 3 (rev 02)
ff:17.7 System peripheral: Intel Corporation Xeon E7 v3/Xeon E5 v3/Core i7 DDRIO (VMSE) 2 & 3 (rev 02)
ff:1e.0 System peripheral: Intel Corporation Xeon E7 v3/Xeon E5 v3/Core i7 Power Control Unit (rev 02)
ff:1e.1 System peripheral: Intel Corporation Xeon E7 v3/Xeon E5 v3/Core i7 Power Control Unit (rev 02)
ff:1e.2 System peripheral: Intel Corporation Xeon E7 v3/Xeon E5 v3/Core i7 Power Control Unit (rev 02)
ff:1e.3 System peripheral: Intel Corporation Xeon E7 v3/Xeon E5 v3/Core i7 Power Control Unit (rev 02)
ff:1e.4 System peripheral: Intel Corporation Xeon E7 v3/Xeon E5 v3/Core i7 Power Control Unit (rev 02)
ff:1f.0 System peripheral: Intel Corporation Xeon E7 v3/Xeon E5 v3/Core i7 VCU (rev 02)
ff:1f.2 System peripheral: Intel Corporation Xeon E7 v3/Xeon E5 v3/Core i7 VCU (rev 02)
\end{lstlisting}
\end{document}